\documentclass[aps,twocolumn,secnumarabic,nobalancelastpage,amsmath,amssymb,
nofootinbib]{revtex4}

\usepackage{graphics}      % standard graphics specifications
\usepackage{graphicx}      % alternative graphics specifications
\usepackage{url}           % for on-line citations
\usepackage{bm}            % special 'bold-math' package
\usepackage{mathrsfs}
\usepackage{float}
\usepackage{siunitx}
\usepackage{tabularx}
\usepackage{booktabs}   % optional, for \toprule etc.
\usepackage{microtype}
\usepackage{booktabs}
\usepackage{siunitx}
\usepackage{subcaption} 
\begin{document}
\title{Temperature Dependence of the Momentum-Resolved Static Spin Susceptibility\\
in a Mott-Proximate Cuprate Model with an antinodal gap\\}
%%%%%%%%%%%%%%%%%%%%%%%%%%%%%%%%%%%%%%%%%%%%%%%%%%%%%%%%%%%%%%%%%%%%%%%%%%%%%%%%%%%
\author         {Keishichiro Tanaka$^1$}
\email          {keishichiro.tanaka@gmail.com\
ORCiD:https://orcid.org/0009-0006-3140-2215}
%\homepage       { }
%\affiliation    { }
\date{\today}

%%%%%%%%%%%%%%%%%%%%%%%%%%%%%%%%%%%%%%%%%%%%%%%%%%%%%%%%%%%%%%%%%%%%%%%%%%%%%%%%%%%
%%%%%%%%%%%%%%%%%%%%%%%%%%%%%%%%%%%%%%%%%%%%%%%%%%%%%%%%%%%%%%%%%%%%%%%%%%%%%%%%%%%
\begin{abstract}
This paper presents the temperature dependence of the static spin susceptibility at $\mathbf{q} = (\pi, \pi)$ and $\mathbf{q} = (\pi, 0)$ in a Mott-proximate cuprate model with an antinodal pseudogap -- a model system for high-temperature superconducting (HTSC) cuprates. 

The results show the spin susceptibility onset temperature tracks the critical temperature ($T_c$) of HTSCs with a comparable scale across the electron filling factor.
Also, as the electron filling decreases and the chemical potential approaches the antinodal van Hove region, the spin susceptibility at $\mathbf{q}=(\pi,\pi)$ and $(\pi,0)$, corresponding to diagonal and axial particle-hole scattering channels, respectively, grows markedly.

It suggests that the emergence of cuprate superconductivity correlates with a suppression of low-energy antinodal spin response and associated particle-hole excitations, which would otherwise dephase $d$-wave pairing, commonly attributed to spin fluctuations.
In this context, the pseudogap partially suppresses antinodal spectral weight near $\omega = 0$, thereby reducing the low-$\omega$ particle-hole phase space.
\end{abstract}
\maketitle
%%%%%%%%%%%%%%%%%%%%%%%%%%%%%%%%%%%%%%%%%%%%%%%%%%%%%%%%%%%%%%%%%%%%%%%%%%%%%%%%%%
%%%%%%%%%%%%%%%%%%%%%%%%%%%%%%%%%%%%%%%%%%%%%%%%%%%%%%%%%%%%%%%%%%%%%%%%%%%%%%%%%%

%%%%%%%%%%%%%%%%%%%%%%%%%%%%%%%%%%%%%%%%%%%%%%%%%%%%%%%%%%%%%%%%%%%%%%%%%%%%%%%%%%
%%%%%%%%%%%%%%%%%%%%%%%%%%%%%%%%%%%%%%%%%%%%%%%%%%%%%%%%%%%%%%%%%%%%%%%%%%%%%%%%%%
%%%%%%%%%%%%%%%%%%%%%%%%%%%%%%%%%%%%%%%%%%%%%%%%%%%%%%%%%%%%%%%%%%%%%%%%%%%%%%%%%%
%%%%%%%%%%%%%%%%%%%%%%%%%%%%%%%%%%%%%%%%%%%%%%%%%%%%%%%%%%%%%%%%%%%%%%%%%%%%%%%%%%

%%%%%%%%%%%%%%%%%%%%%%%%%%%%%%%%%%%%%%%%%%%%%%%%%%%%%%%%%%%%%%%%%%%%%%%%%%%%%%%%%%
%%%%%%%%%%%%%%%%%%%%%%%%%%%%%%%%%%%%%%%%%%%%%%%%%%%%%%%%%%%%%%%%%%%%%%%%%%%%%%%%%%
\section{Introduction}
{\small

We study the temperature dependence of the momentum-resolved, static spin susceptibility at the scattering vectors $\mathbf{q}=(\pi,\pi)$ (zone corner) and $\mathbf{q}=(\pi,0)$ (bond-direction zone boundary), 
which probe distinct antinodal and near-nodal scattering channels in the cuprate Brillouin zone, 
using a cuprate model in the underdoped, Mott-proximate regime with an antinodal pseudogap. 

This work is part of a series of studies to elucidate the superconducting mechanism in high-temperature superconducting (HTSC) cuprates 
~\cite{reference1, reference2, reference3, reference4}.
%Our previous study showed the pseudogap of HTSC is a shift in an excitation of J, antiferromagnetic coupling interaction, due to self-energy effects of the system. 
%Accordingly, we concluded that superconductivity in HTSC cuprates appears under antiferromagnetic spin correlations 
We hypothesize that the onset temperature $T^\ast$ of the spin susceptibility at $\mathbf{q}=(\pi,\pi)$ and $(\pi,0)$ is closely correlated with the superconducting critical temperature $T_c$, 
suggesting a common underlying energy scale
~\cite{reference5, reference6, reference7, reference8, reference9, reference10, reference11}.
This hypothesis is motivated by the following observations:
\begin{itemize}
\item Short-range magnetic correlations associated with the antinodal regions have been reported experimentally~\cite{reference12, reference13, reference14, reference15}.
\item Preliminary calculations indicate that, for a $d$-wave-like gap, the effective activation scale extracted from the spin susceptibility $\chi_{zz}(T)$ becomes comparable to $k_B T_C$ due to contributions from nodal regions and multiple momentum-transfer channels.
\end{itemize}

In this study, first, as a baseline for the discussion, we confirm the uniform static spin susceptibility.
Second, we evaluate the temperature dependence of the real part of the static spin susceptibility
$\chi_{zz}(\mathbf{q},\omega=0;T)$ at $\mathbf{q}=(\pi,\pi)$ and $\mathbf{q}=(\pi,0)$ over the entire Brillouin zone 
using a tight-binding model that includes $k$-dependent gap %$\Delta_{(\mathbf{k})}$  
to extract an onset temperature $T^\ast$ from the Arrhenius-type susceptibility.
% and compare it with $T_c$.

In this analysis, we compute the dressed-bubble susceptibility (the Green's function bubble with self-energy), and assess it using Sommerfeld and Arrhenius form. 
The susceptibility kernel incorporates a $\mathbf{k}$-dependent $d$-wave-like BCS-type gap; in addition, employs a Gaussian energy-window scheme to effectively isolate thermally-activated susceptibility $\chi(\mathbf{q};T)$.
Furthermore, we investigate the relationship between the superconductivity phase and the enhancement of electron-hole transitions due to the van Hove singularity (VHS) 
~\cite{reference16, reference17, reference18, reference19, reference20, reference21, reference22}.

The self-energy $\Sigma$ is obtained from cluster dynamical mean-field theory (CDMFT) on a $2\times2$ cluster and periodized to $\mathbf{k}$-space using the cumulant scheme as shown in Appendix~B ~~\cite{reference19, reference23, reference24}.
The chemical potential $\mu$ is determined at each temperature to reproduce the target density.
Pseudogap amplitudes $\Delta (n)$ at antinodes, as a function of electron density (electron filling factor) $n$, are determined based on past experimental results and previous related research in Appendix~A ~\cite{reference10, reference11}.

This paper is structured as follows. Section 2 explains the theoretical model and tight-binding dispersion used in this study. Section 3 introduces susceptibility formalism. Section 4 presents the results of the calculations and validations. The discussions and conclusions are in Section 5 and Section 6, respectively. 

Throughout this paper, $\mathbf{q}$ (scattering vector), $\omega$ (frequency), and $T$ (temperature) are the arguments of the spin susceptibility $\chi_{zz}(\mathbf{q},\omega;\,T )$. Critical and susceptibility onset temperatures are denoted as $T_c$ and $T^\ast$. $\Delta (n) (\equiv \Delta_{\mathrm{nom}}(n)\equiv\Delta_0(n)$) generally denotes a pseudogap scale at antinodes in $\mathrm{eV}$. In particular, $\Delta_{\mathbf{k}}$ denotes the momentum-dependent $d$-wave gap function with maximum amplitude $\Delta_0$. The quantity $\Delta_{\mathrm{nom}}$ denotes the nominal input gap used in the Arrhenius analysis, while $\Delta_{\mathrm{fit}}$ denotes the effective activation scale extracted from the fits. “Eq.”/“Eqs.” are denoted for Equation(s) and “Fig.”/“Figs.” for Figure(s).

}
%%%%%%%%%%%%%%%%%%%%%%%%%%%%%%%%%%%%%%%%%%%%%%%%%%%%%%%%%%%%%%%%%%%%%%%%%%%%%%%%%%
%%%%%%%%%%%%%%%%%%%%%%%%%%%%%%%%%%%%%%%%%%%%%%%%%%%%%%%%%%%%%%%%%%%%%%%%%%%%%%%%%%

%%%%%%%%%%%%%%%%%%%%%%%%%%%%%%%%%%%%%%%%%%%%%%%%%%%%%%%%%%%%%%%%%%%%%%%%%%%%%%%%%%
%%%%%%%%%%%%%%%%%%%%%%%%%%%%%%%%%%%%%%%%%%%%%%%%%%%%%%%%%%%%%%%%%%%%%%%%%%%%%%%%%%
\section{Theoretical Model and Tight-Binding Dispersion}
{\small

The static spin susceptibility in this study is calculated based on a single-band two-dimensional tight-binding dispersion:
\begin{subequations}
\begin{align}
\varepsilon_{\mathbf{k}}
=
-2t(\cos k_x+\cos k_y)
-4t'\cos k_x\cos k_y ,
\end{align}
where $t$ and $t'$ denote the nearest- and next-nearest-neighbor hopping amplitudes, respectively. 
Typical cuprate parameters in the underdoped regime are $t=0.4~\mathrm{eV}$ and $t'=-0.3\,t$. The lattice constant set to unity.

The corresponding band energy measured from the bare chemical potential is
\begin{align}
\xi_{\mathbf{k}}
=
\varepsilon_{\mathbf{k}}
-
\mu_{\mathrm{bare}} .
\end{align}

A $d$-wave-like pseudogap is introduced as
\begin{align}
\Delta_{\mathbf{k}}
=
\Delta_{0} (\cos k_x - \cos k_y)
\exp\!\left(
-\frac{\xi_{\mathbf{k}}^2}{\sigma_{E}^2}
\right).
\end{align}
where $\Delta_{0}$ is the gap amplitude, and $\sigma_E$ controls the radial (energy) extent of the gap.

To incorporate the pseudogap into the low-energy single-particle spectrum, we employ an effective auxiliary $2\times2$ representation within a Bogoliubov--de Gennes (BdG) framework, which phenomenologically captures the approximately particle--hole symmetric structure of the low-temperature pseudogapped spectrum ~\cite{reference21, reference22}.
\begin{align}
H &= \sum_{\mathbf{k}}
\begin{pmatrix}
\phi^\dagger_{\mathbf{k},1} & \phi^\dagger_{\mathbf{k},2}
\end{pmatrix}
\begin{pmatrix}
\xi_{\mathbf{k}} & \Delta_{\mathbf{k}} \\
\Delta_{\mathbf{k}}^\ast & -\xi_{\mathbf{k}}
\end{pmatrix}
\begin{pmatrix}
\phi_{\mathbf{k},1} \\
\phi_{\mathbf{k},2}
\end{pmatrix}.
%\xi_{\mathbf{k}} &= \varepsilon_{\mathbf{k}} - \mu .
\end{align}

Here, $\phi_{\mathbf{k},1}$ and $\phi_{\mathbf{k},2}$ denote effective
electron-like and hole-like degrees of freedom at momentum $\mathbf{k}$.
They are introduced as an auxiliary two-component basis to generate a
particle--hole symmetric gapped spectrum and do not correspond to a
superconducting Nambu spinor.

The resulting effective single-particle spectrum is
\begin{align}
E_{\mathbf{k}} = \pm \sqrt{\xi_{\mathbf{k}}^2 + |\Delta_{\mathbf{k}}|^2}.
\end{align}
\end{subequations}
%In our evaluation, the low-$T$ suppression is governed by the minimum two-quasiparticle excitation energy
%\[
%E_{\min}(\mathbf q) = \min_{\mathbf k} \big[ E_{\mathbf k} + E_{\mathbf{k+q}} \big],
%\]
%rather than the universal $\Delta_{\mathrm{fit}} \sim 2\Delta_{\mathrm{nom}}$ expected for a momentum-independent isotropic gap.
%This leads to an Arrhenius form with an effective $\Delta_{\mathrm{fit}}$, reflecting the minimum excitation energy selected by the available phase space and momentum structure of the system.

The auxiliary electron--hole basis becomes particularly important when
$|\xi_{\mathbf{k}}| \lesssim |\Delta_{\mathbf{k}}|$, where the gap
substantially modifies the low-energy spectrum.
Although the spectrum consists of quasiparticle branches at
$\pm E_{\mathbf{k}}$, the low-energy excitation scale near the Fermi
surface is primarily governed by $|\Delta_{\mathbf{k}}|$.
This formulation enables the analysis of thermally activated
(Arrhenius-like) behavior in response functions while preserving the
underlying Fermi-surface symmetry.

}
%%%%%%%%%%%%%%%%%%%%%%%%%%%%%%%%%%%%%%%%%%%%%%%%%%%%%%%%%%%%%%%%%%%%%%%%%%%%%%%%%%
%%%%%%%%%%%%%%%%%%%%%%%%%%%%%%%%%%%%%%%%%%%%%%%%%%%%%%%%%%%%%%%%%%%%%%%%%%%%%%%%%%

%%%%%%%%%%%%%%%%%%%%%%%%%%%%%%%%%%%%%%%%%%%%%%%%%%%%%%%%%%%%%%%%%%%%%%%%%%%%%%%%%%
%%%%%%%%%%%%%%%%%%%%%%%%%%%%%%%%%%%%%%%%%%%%%%%%%%%%%%%%%%%%%%%%%%%%%%%%%%%%%%%%%%
\section{Susceptibility Formalism}
{\small

In this section, we summarize the mathematical framework used in this study.
We employ two related formulations to calculate the static spin susceptibility of a gapped cuprate model system: primarily, the dressed-bubble formulation (i.e., the Green's-function bubble including self-energy effects), 
and, for reference, the bare Lindhard formulation without self-energy effects for the uniform static spin susceptibility.
Both describe particle--hole excitations and have closely related formal structures, but only the dressed-bubble formulation explicitly incorporates the system self-energy.

To characterize the temperature dependence of the spin susceptibility, we use two fitting schemes:
a Sommerfeld expansion to describe weak, Pauli-like $T^{2}$ behavior, and
an Arrhenius form to capture thermally activated behavior.

For numerical stability, the dressed-bubble calculation employs a Gaussian energy window, 
which effectively restricts the summation to states near the Fermi surface.

}
%%%%%%%%%%%%%%%%%%%%%%%%%%%%%%%%%%%%%%%%%%%%%%%%%%%%%%%%%%%%%%%%%%%%%%%%%%%%%%%%%%

%%%%%%%%%%%%%%%%%%%%%%%%%%%%%%%%%%%%%%%%%%%%%%%%%%%%%%%%%%%%%%%%%%%%%%%%%%%%%%%%%%
\subsection{Lindhard formalism}
{\small

The Lindhard (bare) susceptibility is obtained from particle–hole excitations and is given by
Eq.~\eqref{eq:group-chi-lindhard}, derived from linear response (see Appendix~C)
\cite{reference16, reference17, reference18, reference19}.

\begin{subequations}
\label{eq:group-chi-lindhard}
\begin{align}
\label{eq:l1}
\chi_{0}(\mathbf q,\omega)
  &= -\sum_{\mathbf k}\,
     \frac{ f\!\big(E_{\mathbf k}\big)-f\!\big(E_{\mathbf{k}+\mathbf q}\big) }
          { \,\omega + \big(E_{\mathbf k}-E_{\mathbf{k}+\mathbf q}\big) + i0^{+} } ,
\\[4pt]
\label{eq:l2}
E_{\mathbf k} &= \sqrt{\xi_{\mathbf k}^{2}+\Delta^{2}},\qquad
\xi_{\mathbf k}=\varepsilon_{\mathbf k}-\mu_{\mathrm{bare}},
\\[4pt]
\label{eq:l3}
f(E) &= \frac{1}{e^{\beta E}+1}\, .
\end{align}
\end{subequations}

Here, $\varepsilon_{\mathbf k}$ is the band dispersion, and
$\Delta$ is a \emph{single-particle} gap applied at both $\mathbf k$ and $\mathbf{k}+\mathbf q$.

}
%%%%%%%%%%%%%%%%%%%%%%%%%%%%%%%%%%%%%%%%%%%%%%%%%%%%%%%%%%%%%%%%%%%%%%%%%%%%%%%%%%

%%%%%%%%%%%%%%%%%%%%%%%%%%%%%%%%%%%%%%%%%%%%%%%%%%%%%%%%%%%%%%%%%%%%%%%%%%%%%%%%%%
\subsection{Dressed-bubble formalism}
{\small
The dressed-bubble susceptibility is given by the particle--hole bubble constructed from two fully dressed normal-state Green’s functions and is expressed as
Eq.~(\ref{eq:group-chi-bubble}) ~\cite{reference16,  reference17, reference18, reference19, reference20, reference21, reference22}.

\begin{subequations}\label{eq:group-chi-bubble}
\begin{align}
\label{eq:bubble_chi_GG}
\chi(\mathbf q,i\Omega_m)
  &= -\frac{1}{\beta}\sum_{\mathbf k,n}
     G(\mathbf k,i\omega_n)\,
     G(\mathbf{k}+\mathbf q,\,i\omega_n+i\Omega_m),\\[4pt]
\label{eq:bubble_freqs}
i\omega_n&=(2n{+}1)\pi/\beta,\qquad i\Omega_m=2m\pi/\beta.
\end{align}
where $i\omega_n$ is the fermionic Matsubara frequency and $i\Omega_m$ is the bosonic Matsubara frequency.

The normal component of the BdG-type Green’s function incorporates a frequency-independent gap $\Delta_{\mathbf k}$ and self-energy $\Sigma$ (treated within a static, real-part approximation), 
and is expressed in terms of the quasiparticle energy defined in Eq.~\eqref{eq:E_static}:
\begin{align}
\label{eq:E_static}
E_{\mathbf k}&=\sqrt{\xi_{\mathbf k}^2+|\Delta_{\mathbf k}|^2},\\[4pt]
\xi_{\mathbf k} &= \varepsilon_{\mathbf k} + \mathrm{Re}\,\Sigma(\mathbf k,0)
                 - \mu_{\mathrm{eff}},\\[4pt]
\mu_{\mathrm{eff}} &= \mu_{\mathrm{bare}} + \mathrm{Re}\,\Sigma(\mathbf k_F,0), 
\label{eq:bubble_mu_eff}\\[4pt]
\label{eq:G_static}
G(\mathbf k,i\omega_n)
&= -\,\frac{i\omega_n+\xi_{\mathbf k}}{\omega_n^2+E_{\mathbf k}^2}\,.
\end{align}
\end{subequations}

Here $\mu_{\mathrm{eff}}$ is the effective chemical potential, 
reported simply as $\mu$ in figures and tables.  
We evaluate the static spin susceptibility from the $m=0$ bosonic 
Matsubara component in the static limit, 
$\chi(\mathbf q,0)\equiv\chi(\mathbf q,i\Omega_m{=}0)$, 
which coincides with the retarded susceptibility $\chi^{R}(\mathbf q,\omega)$ at $\omega=0$, 
provided the response function is regular at zero frequency.

}
%%%%%%%%%%%%%%%%%%%%%%%%%%%%%%%%%%%%%%%%%%%%%%%%%%%%%%%%%%%%%%%%%%%%%%%%%%%%%%%%%%
%%%%%%%%%%%%%%%%%%%%%%%%%%%%%%%%%%%%%%%%%%%%%%%%%%%%%%%%%%%%%%%%%%%%%%%%%%%%%%%%%%

%%%%%%%%%%%%%%%%%%%%%%%%%%%%%%%%%%%%%%%%%%%%%%%%%%%%%%%%%%%%%%%%%%%%%%%%%%%%%%%%%%
%%%%%%%%%%%%%%%%%%%%%%%%%%%%%%%%%%%%%%%%%%%%%%%%%%%%%%%%%%%%%%%%%%%%%%%%%%%%%%%%%%
\section{Results}
{\small

We first compute the uniform static spin susceptibility using both the dressed bubble and the Lindhard bare formulation and assess it via the Sommerfeld form.

Next, using the dressed-bubble formulation, we compute the temperature dependence of the static spin susceptibility $\chi_{zz}(\mathbf{q},\omega=0;T)$ with a $d$-wave-type gap ($\Delta>0$) at $\mathbf{q}=(\pi,\pi)$ and $\mathbf{q}=(\pi,0)$, and assess the thermally activated behavior using Arrhenius fits.

Then, we illustrate the $\mathbf{k}$-resolved contributions to the spin susceptibility %through $I_{\rm abs}$ 
and examine the influence of the van Hove singularity (VHS), as reflected in the underlying joint density-of-states (JDOS) structure.

For the spin susceptibility calculations, a set of 
\[
\Delta_{\mathrm{nom}}(n) = 0.010,\ 0.021,\ 0.034,\ 0.039,\ 0.051,\ 0.054~\mathrm{eV}
\]
is used for
\[
n = 0.95,\ 0.90,\ 0.85,\ 0.80,\ 0.70,\ 0.60,
\]
as obtained from previous results shown in Appendix A.
The effective chemical potentials including self-energy are set as 
\[
\mu_{\mathrm{}}(4\mathrm{K}) =1.159,\,1.118,\,1.075,\,1.036,\,0.970,\ 0.925~\mathrm{eV}
\]
respectively.

Unless stated otherwise, we report the real part of $\chi_{zz}$ computed as a full Brillouin-zone sum using an energy-window scheme, which improves the stability of the thermally activated fits and allows extraction of the peak temperature $T^{*} (\equiv T_{\mathrm{peak}})$ of $\chi_{zz}(\mathbf{q}, \omega=0; T)$ for $\Delta > 0$. 

The Gaussian energy-window parameter $\Lambda$ is set to $1.25\,\Delta_{\mathrm{nom}}(n)$, and the radial width $\sigma_E$ of the $d$-wave-type gap to $0.25\,\Delta_{\mathrm{nom}}(n)$, corresponding to a narrower Gaussian profile around the Fermi surface than the broader choice $\sigma_E=\Delta_{\mathrm{nom}}(n)$. These values were determined from a short parameter sweep initialized at $\Lambda,\sigma_E=\Delta_{\mathrm{nom}}(n)$.

At each temperature, the chemical potential $\mu(T)$ is determined to match the target density using the Fermi–Dirac distribution and the tight-binding dispersion. 
A density of $n=1.0$ corresponds to one electron per site, i.e., half-filling.

The Brillouin-zone average of the real part of the self-energy at the lowest Matsubara frequency is
\[
\langle \mathrm{Re}\,\Sigma(i\omega_0)\rangle_{\mathrm{BZ}} = 1.51~\mathrm{eV}.
\]
This converged metallic solution is obtained from CDMFT calculations with parameters $U = 8.0\,t$, $\beta = 100/t$, and $t = 0.4~\mathrm{eV}$, using a fixed chemical potential $\mu = 3.2\,t$ ($\mu = 4.0\,t$ corresponds to half-filling). The same metallic state is used in all subsequent susceptibility calculations as shown in Appendix~B.

All susceptibilities are normalized per site and expressed in units of $\mathrm{eV}^{-1}$ per unit cell. To convert to the physical spin susceptibility, multiply by $(g\mu_{\mathrm{B}})^{2}/4$; for $g=2$, this reduces to $\mu_{\mathrm{B}}^{2}$, where $\mu_{\mathrm{B}}$ is the Bohr magneton.

}
%%%%%%%%%%%%%%%%%%%%%%%%%%%%%%%%%%%%%%%%%%%%%%%%%%%%%%%%%%%%%%%%%%%%%%%%%%%%%%%%%%

%%%%%%%%%%%%%%%%%%%%%%%%%%%%%%%%%%%%%%%%%%%%%%%%%%%%%%%%%%%%%%%%%%%%%%%%%%%%%%%%%%
\subsection{Fitting procedures}
{\small

The Sommerfeld form in Eq.~(\ref{eq:fit1}), a low-$T$ asymptotic expansion, is employed to model the weak temperature dependence of $\chi(T)$, and is fit using ordinary least squares. The fit is restricted to $T = 11.49$–$33.95~\mathrm{K}$ (the first six points with $T \ge 10~\mathrm{K}$) to focus on the low-$T$ regime, guided by the heuristic $k_{\mathrm B}T / E_F \lesssim 0.05$ and by residual diagnostics.

The Arrhenius-type forms in Eqs.~(\ref{eq:fit2})–(\ref{eq:fit3}) are used to model thermally activated behavior of $\chi(T)$, and are fit by nonlinear least squares over the selected temperature window. These form require a temperature interval in which $\ln\chi$ is approximately linear in $1/T$.

We report parameter estimates with 95\% confidence intervals (CIs) and the coefficient of determination $R^2$ (Eq.~(\ref{eq:fit_r2})).
Here, SSE is the sum of squared residuals between the data and the fitting function, while SST is the total sum of squares relative to the mean value of the data.

\begin{subequations}\label{eq:group-fit}
\begin{align}
\label{eq:fit1}
\chi(T) &= a + b\,(k_{\mathrm B}T)^2,\\
\label{eq:fit2}
\chi(T) &= c_0 + A\, \exp\!\big[-\Delta_{\text{fit}}/(k_{\mathrm B}T)\big],\\
\label{eq:fit3}
\chi(T) &= c_0 + b'\,(k_{\mathrm B}T)^2 + A\, \exp\!\big[-\Delta_{\text{fit}}/(k_{\mathrm B}T)\big],\\
\label{eq:fit_r2}
R^2 &= 1 - \frac{\mathrm{SSE}}{\mathrm{SST}}.
\end{align}
\end{subequations}

Here, $a$ and $c_{0}$ are baseline offsets. The parameter $b$ characterizes a weak quadratic $T^{2}$ dependence, while $b'$ represents a non-activated contribution. The parameter $A$ is the amplitude of the activated contribution. 
The parameter $\Delta_{\text{nom}}$ denotes the nominal (input) gap per quasiparticle, while $\Delta_{\text{fit}}$ is an effective activation energy extracted from the Arrhenius fit to $\chi(T)$.
Physically, $\Delta_{\mathrm{fit}}$ characterizes the low-temperature suppression scale of $\chi_{zz}(\mathbf q,T)$, rather than the imposed gap amplitude itself.

$T_{\mathrm{peak}}$ represents the crossover where thermally activated excitations are sufficiently populated to maximize the response, before phase-space saturation reduces the spin susceptibility.
$T_{\mathrm{peak}}$ is determined as the temperature at which $\chi(T)$ attains a local maximum:
\[
T_{\mathrm{peak}} = \operatorname*{arg\,max}_T \chi(T).
\]

$T_{\mathrm{slope}}$ denotes the temperature at which $\chi(T)$ exhibits the steepest increase, identified by the maximum of its second derivative (restricted to $T \ge 10~\mathrm{K}$):
\[
T_{\mathrm{slope}} = \operatorname*{arg\,max}_{T \ge 10~\mathrm{K}} \frac{d^{2}\tilde{\chi}}{dT^{2}}(T),
\qquad
\tilde{\chi}(T) = S[\chi(T)].
\]
Here, $S$ denotes a smoothing operation. In practice, $d\chi/dT$ is first smoothed using a 5-point moving average prior to evaluating the second derivative.

In this $d$-wave case, where the onset of thermally activated behavior is broadened by nodal and momentum-dependent contributions, the three-parameter fit (Eq.~(\ref{eq:fit2})) provides a better description than the four-parameter fit (Eq.~(\ref{eq:fit3})), likely because the broadened activated response reduces the relative importance of the additional non-activated term.

Accordingly, $T_{\mathrm{peak}}$ is used as a practical proxy for the characteristic temperature scale $T^*$ (or $T_{\mathrm{onset}}$).

}
%%%%%%%%%%%%%%%%%%%%%%%%%%%%%%%%%%%%%%%%%%%%%%%%%%%%%%%%%%%%%%%%%%%%%%%%%%%%%%%%%%

%%%%%%%%%%%%%%%%%%%%%%%%%%%%%%%%%%%%%%%%%%%%%%%%%%%%%%%%%%%%%%%%%%%%%%%%%%%%%%%%%%
\subsection{Uniform susceptibility}
{\small

Fig.~\ref{fig:chi_uniform_q0} shows the temperature dependence of the uniform static spin susceptibility $\chi_{zz}(\mathbf{q}=0,\omega=0;T)$ at $\Delta=0$.
The uniform susceptibilities are calculated using both the dressed-bubble and, as a cross-check, the bare Lindhard formulation (Eq.~ (\ref{eq:group-chi-lindhard}) and (\ref{eq:group-chi-bubble})).
The temperature dependence is then assessed using Sommerfeld form in Eq.~(\ref{eq:fit1}).  

Both Sommerfeld fits indicate a Pauli-like temperature dependence with a weak quadratic ($T^{2}$) correction, consistent with a finite but modest quasiparticle DOS $N^{*}(E_{F})$.
 
The chemical potentials are set as $\mu(4\mathrm{K}) =-0.265~\mathrm{eV}$ for the Lindhard, whereas set as $\mu(4\mathrm{K})=1.207~\mathrm{eV}$ for the dressed-bubble, both to match electron density $n=1.0$ (metallic case).

The fits yield $a = 0.398$, $b = -1.07 \times10^{4} \mathrm{K}^{-2}$, $R^{2} = 0.849$ for the Lindhard susceptibility, 
and $a = 0.981$, $b = -4.83 \times 10^{4}~\mathrm{K}^{-2}$, and $R^{2} = 0.893$ for the dressed-bubble susceptibility (Table~\ref{tab:sommerfeld_q0}). 

\begin{figure}[H]
    \centering
    \includegraphics[width=0.35\textwidth]{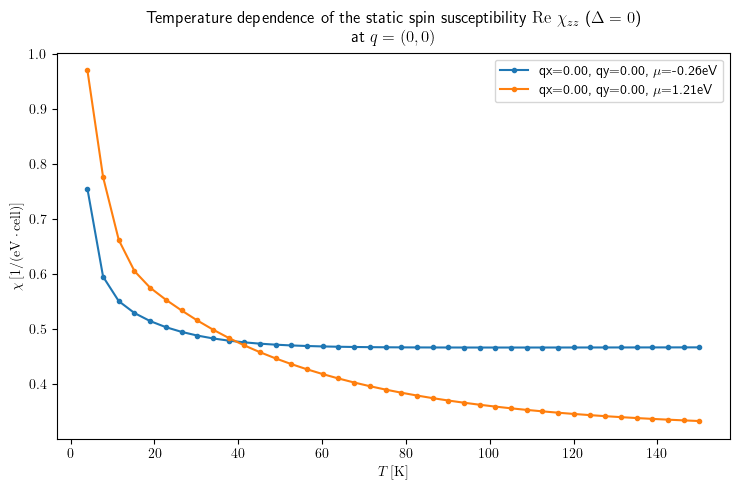}
    \caption{\footnotesize
    Temperature dependence of $\chi(\mathbf{q}=0;T)$ at $\Delta=0$,
    evaluated using the Lindhard bare (blue and flatter curve) and the dressed-bubble susceptibility (orange). 
    The chemical potentials are $\mu(4\mathrm{K})=-0.26~\mathrm{eV}$ and $1.21~\mathrm{eV}$, corresponding to $n=1.0$ (metallic case).}
    \label{fig:chi_uniform_q0}
\end{figure}

\begin{table}[h]
\centering
\caption{\footnotesize
Sommerfeld fits for the uniform static spin susceptibility
$\chi(\mathbf{q}=0;T)$ over $T=11.49$–$33.95~\mathrm{K}$ 
(first six points $\ge10~\mathrm{K}$) at $\mathbf{q}=(0,0)$ and $\Delta=0$.
Parameters $a$ and $b$ are obtained from ordinary least-squares fits of
$\chi=a+bT^{2}$, with 95\% confidence intervals (CIs) shown in brackets.}
\label{tab:sommerfeld_q0}
{\footnotesize
\begin{tabular}{
l
S[table-format=1.3]
S[table-format=2.2]
S[table-format=1.3]
}
\toprule
Model & {$a$ [95\% CI]} & {$b$ [95\% CI] ($10^{4}\,\mathrm{K}^{-2}$)} & {$R^{2}$}\\
\midrule
\text{Lindhard (bare)} & {0.398 [0.373,\,0.424]} & {-1.07 [-1.69,\, -0.441]} & 0.849 \\
\text{Bubble}          & {0.981 [0.887,\,1.075]} & {-4.83 [-7.15,\, -2.52]} & 0.893 \\
\bottomrule
\end{tabular}
}
\end{table}

\noindent\textit{Note.}
The uniform static spin susceptibility is
\setlength{\abovedisplayskip}{4pt}
\setlength{\belowdisplayskip}{4pt}
\[
\chi_u(T) \equiv \chi(\mathbf{q}=0,\omega=0;T)
= \left.\frac{\partial M}{\partial h}\right|_{h\to 0} .
\]
i.e., the response to a spatially uniform field. In a Fermi-liquid it primary reflects the quasiparticle DOS at $E_{F}$ and thus provides the Pauli-like baseline
%(exchange interactions give a modest multiplicative enhancement)”
%In the Fermi-liquid regime, a Pauli-like uniform susceptibility is governed by the \emph{interacting} single-particle density of states at the Fermi level, so we write
%$\chi_u \propto N^{\ast}(E_F) + \mathcal{O}(T^2)$
%when $N^{\ast}(E)$ is smooth near $E_F$; this indicates itinerant quasiparticles with finite $N^{\ast}(E_F)$ and does not imply an absence of interactions. 
 }
%%%%%%%%%%%%%%%%%%%%%%%%%%%%%%%%%%%%%%%%%%%%%%%%%%%%%%%%%%%%%%%%%%%%%%%%%%%%%%%%%%

%%%%%%%%%%%%%%%%%%%%%%%%%%%%%%%%%%%%%%%%%%%%%%%%%%%%%%%%%%%%%%%%%%%%%%%%%%%%%%%%%%
\subsection{Temperature dependence of $\chi_{zz}$}
{\small

Figs.~\ref{fig:chi_static_pipi}\subref{fig:chi_pipi_main}--\ref{fig:chi_static_pizero}\subref{fig:chi_pizero_vhs} show the temperature dependence of the static spin susceptibility $\chi_{zz}(\mathbf{q},\omega=0;T)$ with a $d$-wave-type gap ($\Delta_{\mathrm{nom}}>0$), calculated using the dressed-bubble formulation (Eq.~(\ref{eq:group-chi-bubble})), at $\mathbf{q}=(\pi,\pi)$ and $\mathbf{q}=(\pi,0)$ over the density range $n=0.6$--$0.95$.

At $\mathbf{q}=(\pi,0)$, the response is closely related to incommensurate contributions at $\mathbf{q}=(\pi,\pm\delta)$, which emerge in the present model from the momentum-dependent gap and the associated phase-space structure near the Fermi surface.

For $\Delta_{\mathrm{nom}}>0$, the spin susceptibilities at both $\mathbf{q}=(\pi,\pi)$ and $\mathbf{q}=(\pi,0)$ exhibit an overall Arrhenius-type temperature dependence. Over the density range $n=0.8$--$0.95$, this behavior is quantified using the Arrhenius form (Eq.~(\ref{eq:fit2})). As $n$ decreases, both $\Delta_{\mathrm{nom}}$ and the extracted activation scale $\Delta_{\mathrm{fit}}$ increase systematically, demonstrating a consistent correspondence between the imposed gap scale and the Arrhenius behavior of the susceptibility, as summarized in Table~\ref{tab:static_arrhenius}.

The amplitude at $\mathbf{q} = (\pi,\pi)$ increases monotonically with decreasing $n$, shows a pronounced enhancement at $n = 0.7$, and then decreases at $n = 0.6$. At $\mathbf{q} = (\pi,0)$, the amplitude remains nearly constant from $n = 0.95$ to $0.8$, increases at $n = 0.7$, and grows significantly at $n = 0.6$.

The characteristic temperatures (defined here as the peak temperatures of $\chi_{zz}$) are 
$T^{*} = 33.7,\ 65.9,\ 107.9,\ 100.5~\mathrm{K}$ at $\mathbf{q} = (\pi,\pi)$ and 
$T^{*} = 9.2,\ 40.7,\ 84.9,\ 101.7~\mathrm{K}$ at $\mathbf{q} = (\pi,0)$ 
for $n=0.95,\,0.9,\,0.85,\,0.8$, respectively (Table~\ref{tab:static_arrhenius}). 
%These values scale approximately as $T^{*} \sim \Delta_{\mathrm{nom}}/k_{\mathrm B}$, indicating that larger gap amplitudes shift the onset of thermally activated particle–hole excitations to higher temperatures. Consequently, $\chi(\mathbf{q}, T)$ departs from its low-temperature baseline at higher $T$.

%At lower densities, $n = 0.70$ and $0.60$, corresponding to $\mu(4\,\mathrm{K}) = 0.970~\mathrm{eV}$ and $0.925~\mathrm{eV}$, respectively, the influence of the van Hove singularity (VHS) at the antinodes becomes apparent, as shown in Figs.~\ref{fig:chi_static_pipi}\subref{fig:chi_pipi_vhs} and \ref{fig:chi_static_pizero}\subref{fig:chi_pizero_vhs}.

For each $n$, we perform a local $q$-scan over $q_y \in [0,0.1\pi]$ around $(\pi,\pi)$ and $(\pi,0)$, and select the branch nearest to the target scattering vector that exhibits the smoothest peak structure.
Curves showing irregular or discontinuous behavior are discarded as numerical artifacts.

We fit the spin susceptibility using a three-parameter Arrhenius form,
\[
\chi(\mathbf{q},T)
= c_{0}
+ A\,\exp\!\big[-\Delta_{\mathrm{fit}}/(k_{\mathrm B}T)\big],
\]
over the range $T \in [4,150]~\mathrm{K}$. The fitting parameters $\Delta_{\mathrm{fit}}$, $c_{0}$, and $A$ (with $A>0$), together with the coefficient of determination $R^{2}$ and 95\% confidence intervals, are reported.

%This form provides a stable description primarily for larger gap values, whereas smaller-gap cases exhibit a less well-defined activated regime due to nodal quasiparticle contributions.

In contrast to the $(\pi,0)$ channel, where $\Delta_{\mathrm{fit}} < \Delta_{\mathrm{nom}}$, the spin susceptibility at $(\pi,\pi)$ shows more robust Arrhenius behavior, with $\Delta_{\mathrm{fit}} \sim \Delta_{\mathrm{nom}}$ and consistently higher $R^{2}$ values.% reflecting the relatively stronger antinodal contribution and reduced influence of low-energy nodal excitations.

Overall, larger $\Delta_{\mathrm{nom}}$ yields more pronounced Arrhenius behavior (higher $R^{2}$). %reflecting the suppression of low-energy contributions and a clearer separation of the activation scale.

Minor deviations from ideal Arrhenius behavior may arise from finite-temperature Fermi-window effects, the limited fitting range, and artificial spectral broadening.

%%%%%
% Figures q=(pi,pi)
%%%%%
\begin{figure*}[t]
\centering

\begin{subfigure}{0.35\textwidth}
    \includegraphics[width=\linewidth]{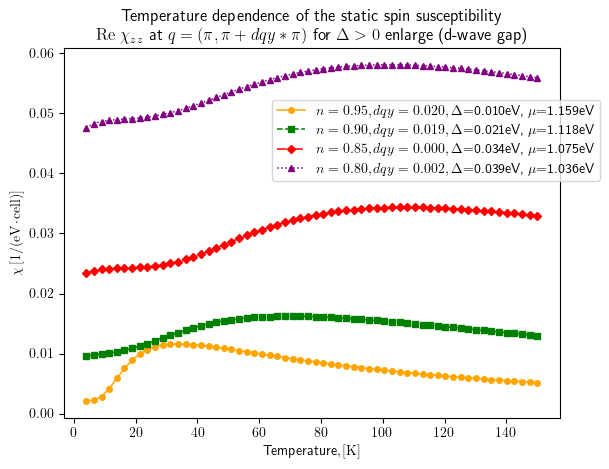}
    \caption{\footnotesize
    $\chi(\mathbf{q}, T)$ at $\mathbf{q}=(\pi,\pi)$.
    Curves (top to bottom at $150~\mathrm{K}$): $n = 0.80, 0.85, 0.90, 0.95$.}
    \label{fig:chi_pipi_main}
\end{subfigure}
\hfill
\begin{subfigure}{0.35\textwidth}
    \includegraphics[width=\linewidth]{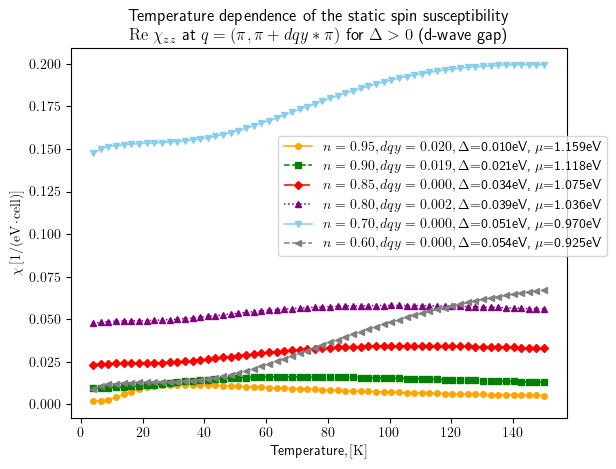}
    \caption{\footnotesize
    $\chi(\mathbf{q}, T)$ at $\mathbf{q}=(\pi,\pi)$, including the VHS regime.
    Curves (top to bottom at $150~\mathrm{K}$): $n = 0.70, 0.60, 0.80, 0.85, 0.90, 0.95$.}
    \label{fig:chi_pipi_vhs}
\end{subfigure}

\caption{\footnotesize
Temperature dependence of the dressed-bubble susceptibility with a $d$-wave gap at $\mathbf{q}=(\pi,\pi)$.
(a) Main density range.
(b) Extended range including the VHS regime.
Parameters: density $n$, gap $\Delta$ (eV), and chemical potential $\mu(4~\mathrm{K})$ (eV).
}
\label{fig:chi_static_pipi}
\end{figure*}

%%%%%
%  Figures q=(pi,0)
%%%%%
\begin{figure*}[t]
\centering

\begin{subfigure}{0.35\textwidth}
    \includegraphics[width=\linewidth]{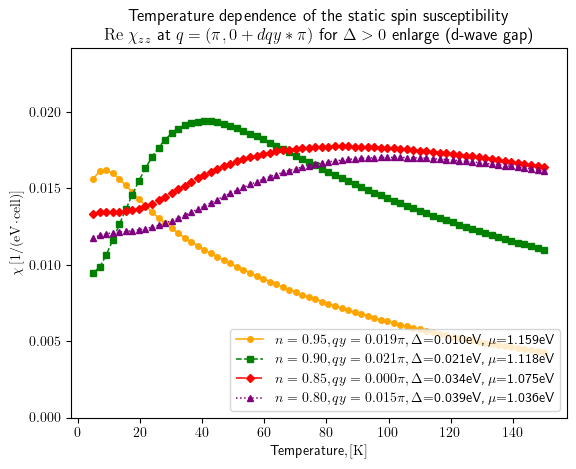}
    \caption{\footnotesize
    $\chi(\mathbf{q}, T)$ at $\mathbf{q}=(\pi,0)$.
    Curves (top to bottom at $150~\mathrm{K}$): $n = 0.85, 0.80, 0.90, 0.95$.}
    \label{fig:chi_pizero_main}
\end{subfigure}
\hfill
\begin{subfigure}{0.35\textwidth}
    \includegraphics[width=\linewidth]{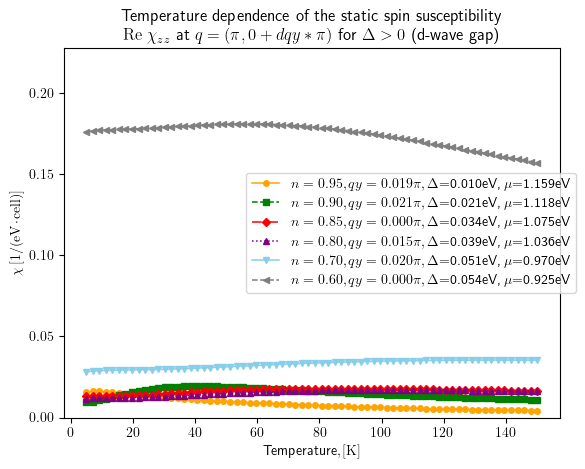}
    \caption{\footnotesize
    $\chi(\mathbf{q}, T)$ at $\mathbf{q}=(\pi,0)$, including the VHS regime.
    Curves (top to bottom at $150~\mathrm{K}$): $n = 0.60, 0.70, 0.85, 0.80, 0.90, 0.95$.}
    \label{fig:chi_pizero_vhs}
\end{subfigure}

\caption{\footnotesize
Temperature dependence of the dressed-bubble susceptibility with a $d$-wave gap at $\mathbf{q}=(\pi,0)$.
(a) Main density range.
(b) Extended range including the VHS regime.
Parameters: density $n$, gap $\Delta$ (eV), and chemical potential $\mu(4~\mathrm{K})$ (eV).
}
\label{fig:chi_static_pizero}
\end{figure*}

%%%%%
% Tables
%%%%%
\begin{table}[h]
\centering
\caption{\footnotesize
Characteristics of the temperature dependence of static spin susceptibility at $\mathbf{q} = (\pi, \pi)$ and $\mathbf{q} = (\pi,  0)$. 
:$T_{slope}$, $T_{peak}(=T^{*})$, $\Delta\chi$(observable change), and the three parameter Arrhenius fits with diagnostic scale comparison. 
}
\label{tab:static_arrhenius}
%{\footnotesize
{\scriptsize

% ---------- (pi,pi) ----------
\textbf{$\mathbf q=(\pi,\pi)$}

\begin{tabular}{
S[table-format=1.2,round-mode=places, round-precision=2]
S[table-format=2.1, round-mode=places, round-precision=1]
S[table-format=3.1, round-mode=places, round-precision=1]
}
\toprule
{$n$}&
{$T_{\text{slope}}$} &
{$T_{\text{peak}}$} \\
\midrule
0.95 & 13.9 & 33.7 \\
0.90 & 31.2 & 65.9 \\
0.85 & 48.5 & 107.9 \\
0.80 & 46.1 & 100.5 \\
\bottomrule
\end{tabular}

\begin{tabular}{
S[table-format=1.3]
S[table-format=1.3]
S[table-format=1.2e-2, round-mode=figures, round-precision=3]
S[table-format=1.2e-2, round-mode=figures, round-precision=3]
S[table-format=1.2e-2, round-mode=figures, round-precision=3]
S[table-format=1.3]
}
\toprule
{$\Delta_{\text{nom}}$} &
{$R^2$} &
{$c_0$} &
{$A$} &
{$\Delta_{\text{fit}}$ (eV)} &
{$\Delta\chi$} \\
\midrule
0.010 &  0.97  & 1.05e-2 & 1.60e-3 & 8.50e-3 & 0.00946 \\
0.021 &  0.995 & 1.20e-2 & 4.80e-3 & 1.85e-2 & 0.00664 \\
0.034 &  0.999 & 2.05e-2 & 9.80e-3 & 3.10e-2 & 0.0110  \\
0.039 &  0.999 & 4.10e-2 & 1.05e-2 & 3.60e-2 & 0.0105  \\
\bottomrule
\end{tabular}

~\\
% ---------- (pi,0) ----------
\textbf{$\mathbf q=(\pi,0)$}

\begin{tabular}{
S[table-format=1.2,round-mode=places, round-precision=2]
S[table-format=2.1, round-mode=places, round-precision=1]
S[table-format=3.1, round-mode=places, round-precision=1]
}
\toprule
{$n$}&
{$T_{\text{slope}}$} &
{$T_{\text{peak}}$} \\
\midrule
0.95 & {N/A} & 9.20 \\
0.90 & 13.4  & 40.7 \\
0.85 & 32.3  & 84.9 \\
0.80 & 42.8  & 101.7 \\
\bottomrule
\end{tabular}

\begin{tabular}{
S[table-format=1.3]
S[table-format=1.3]
S[table-format=1.2e-2, round-mode=figures, round-precision=3]
S[table-format=1.2e-2, round-mode=figures, round-precision=3]
S[table-format=1.2e-2, round-mode=figures, round-precision=3]
S[table-format=1.3]
}
\toprule
{$\Delta_{\text{nom}}$} &
{$R^2$} &
{$c_0$} &
{$A$} &
{$\Delta_{\text{fit}}$ (eV)} &
{$\Delta\chi$} \\
\midrule
0.010 & 0.91  & 1.55e-2 & 6.50e-4 & 4.20e-3 & 0.0108 \\
0.021 & 0.95  & 1.00e-2 & 3.80e-3 & 1.30e-2 & 0.00994 \\
0.034 & 0.97  & 1.28e-2 & 6.20e-3 & 2.20e-2 & 0.00440 \\
0.039 & 0.98  & 1.14e-2 & 7.00e-3 & 2.60e-2 & 0.00529 \\
\bottomrule
\end{tabular}

}
\end{table}

}
%%%%%%%%%%%%%%%%%%%%%%%%%%%%%%%%%%%%%%%%%%%%%%%%%%%%%%%%%%%%%%%%%%%%%%%%%%%%%%%%%%

%%%%%%%%%%%%%%%%%%%%%%%%%%%%%%%%%%%%%%%%%%%%%%%%%%%%%%%%%%%%%%%%%%%%%%%%%%%%%%%%%%
\subsection{K-dependent contributions}
\label{subsec:k-deps}
{\small

We define the $\mathbf{k}$-resolved intensity $I_{\mathbf{k}}(\mathbf{q})$ as the contribution of momentum $\mathbf{k}$ to the bubble susceptibility at fixed momentum transfer $\mathbf{q}$, such that
\begin{equation}
\label{eq:k_intensity}
\chi(\mathbf{q}) = \sum_{\mathbf{k}} I_{\mathbf{k}}(\mathbf{q}).
\end{equation}

We use an energy window $\Lambda = 1.5\,\Delta_{\mathrm{nom}}(n)$ for $\mathbf{q}=(\pi,\pi)$ and $\Lambda = 2.0\,\Delta_{\mathrm{nom}}(n)$ for $\mathbf{q}=(\pi,0)$ in order to enhance the spectral weight in the low-energy region $\sim \Delta_{\mathrm{nom}}(n)$.

Figs.~\ref{fig:k_resolved_combined_pipi}\subref{fig:bz_pipi} and \ref{fig:k_resolved_combined_pizero}\subref{fig:bz_pizero} show $I_{\mathbf{k}}(\mathbf{q})$ in the Brillouin zone for $n=0.8$, $\Delta_{\mathrm{nom}}(n)=0.039~\mathrm{eV}$, and $\mu=1.0352~\mathrm{eV}$ at $T=4~\mathrm{K}$ for $\mathbf{q}=(\pi,\pi)$ and $\mathbf{q}=(\pi,0)$, respectively. The dominant scattering points are connected by straight lines.

The former exhibits the edges of a diamond-shaped structure characteristic of $\mathbf q=(\pi,\pi)$ scattering. In contrast, the latter is dominated by scattering between near-nodal regions, and shows an apparent C2 symmetry arising from the single-$\mathbf q$ selection at $\mathbf q=(\pi,\delta)$.

To recover the $C_4$ symmetry observed experimentally in the axial response, the symmetry-related momentum transfers $\mathbf q=(\pi,\pm\delta)$ and $(\pm\delta,\pi)$ must be included, as shown in Fig.~\ref{fig:k_resolved_combined_others}\subref{fig:bz_pizero_c4}. The corresponding symmetrized intensity is given by
\[
I_{\mathbf{k}}^{\mathrm{sym}}
=
I_{\mathbf{k}}(\pi,\delta)
+
I_{\mathbf{k}}(\pi,-\delta)
+
I_{\mathbf{k}}(\delta,\pi)
+
I_{\mathbf{k}}(-\delta,\pi).
\]

Figs.~\ref{fig:k_resolved_combined_pipi}\subref{fig:top1_pipi} and \ref{fig:k_resolved_combined_pizero}\subref{fig:top1_pizero} show the top-1\% $\mathbf{k}$-resolved contributions for $n=0.8$ at $\mathbf{q}=(\pi,\pi)$ and $\mathbf{q}=(\pi,0)$, respectively, at 4~K and 100~K. The upper panels show the contributions projected onto the $d$-wave coordinate, while the lower panels show their distribution along the Fermi surface. 

At $\mathbf{q}=(\pi,\pi)$, the dominant contributions arise near the antinodal regions. The antinodal contribution becomes stronger from 4~K to 100~K in Fig.~\ref{fig:k_resolved_combined_pipi}\subref{fig:top1_pipi}.
At $\mathbf{q}=(\pi,0)$, the dominant contributions arise near near-nodal regions. In this case as well, the antinodal contribution becomes stronger from 4~K to 100~K, although the corresponding increase in intensity is visually subtle in Fig.~\ref{fig:k_resolved_combined_pizero}\subref{fig:top1_pizero}. 

In the upper panels, $|\cos k_x - \cos k_y| = 2$ corresponds to the antinodal point $\mathbf{k}=(\pi,0)$. In the lower panels, the black line represents the Fermi surface defined by $\xi_{\mathbf{k}}=0$. Some points appear slightly off the FS because they satisfy the finite energy-window condition $|\xi_{\mathbf{k}}| \lesssim \Lambda$ at $\mathbf{q}=(\pi,0)$.

For $\mathbf{q}=(\pi,0)$, the ring-shaped pattern around the M point is an artifact of the cumulant periodization procedure, whereas the corresponding structure around the X point represents a physical scattering contribution.

Fig.~\ref{fig:k_resolved_combined_others}\subref{fig:top1_vhs} shows, for $n=0.7$ ($\Delta_{\mathrm{nom}}(n)=0.051~\mathrm{eV}$, $\mu=0.970~\mathrm{eV}$), the top-1\% $\mathbf{k}$-resolved contributions projected onto the $d$-wave coordinate and the Fermi surface at $\mathbf{q}=(\pi,\pi)$ and $\mathbf{q}=(\pi,0)$ (C2) at 4~K. In both cases, the influence of the van Hove singularity (VHS) at the antinodes becomes visible in the intensity distribution.

\begin{figure*}[t]
\centering

\begin{subfigure}{0.35\textwidth}
    \includegraphics[width=\linewidth]{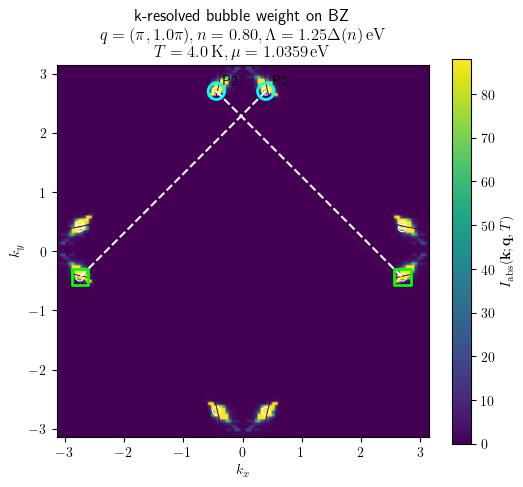}
    \caption{\footnotesize
    $I_{\mathbf{k}}(\mathbf{q})$ at $\mathbf{q}=(\pi,\pi)$.}
    \label{fig:bz_pipi}
\end{subfigure}
\hfill
\begin{subfigure}{0.45\textwidth}
    \includegraphics[width=\linewidth]{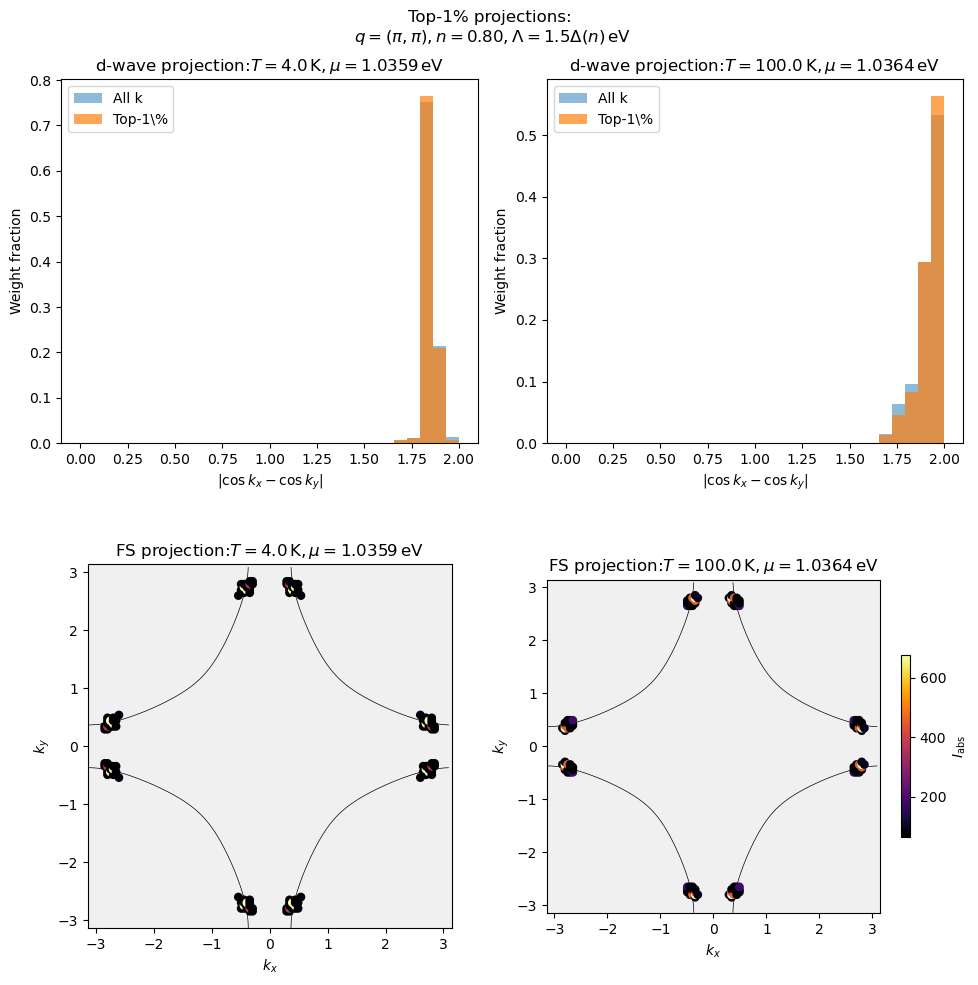}
    \caption{\footnotesize
    Top-1\% $I_{\mathbf{k}}(\mathbf{q})$ at $\mathbf{q}=(\pi,\pi)$.}
    \label{fig:top1_pipi}
\end{subfigure}

\caption{\footnotesize
$\mathbf{k}$-resolved contributions to the static spin susceptibility intensity $I_{\mathbf{k};\mathbf{q}}$ and the corresponding top-1\% weighted distributions for $\mathbf{q}=(\pi,\pi)$.
(a) Brillouin-zone intensity map at $n=0.8$ and $T=4$ K, with a dominant pair of scattering partners connected by a straight line.
(b) Corresponding top-1\% contributions projected onto the $d$-wave coordinate and the Fermi surface at $T=4$ K and $100$ K.
Parameters: $\Delta_{\mathrm{nom}}(n)=0.039$ eV and $\mu=1.0352$ eV for $n=0.8$.
}
\label{fig:k_resolved_combined_pipi}
\end{figure*}

\begin{figure*}[t]
\centering

\begin{subfigure}{0.35\textwidth}
    \includegraphics[width=\linewidth]{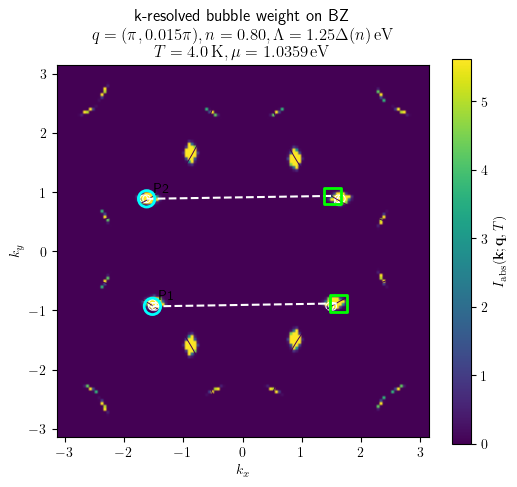}
    \caption{\footnotesize
    $I_{\mathbf{k}}(\mathbf{q})$ at $\mathbf{q}=(\pi, 0)$.}
    \label{fig:bz_pizero}
\end{subfigure}
\hfill
\begin{subfigure}{0.35\textwidth}
    \includegraphics[width=\linewidth]{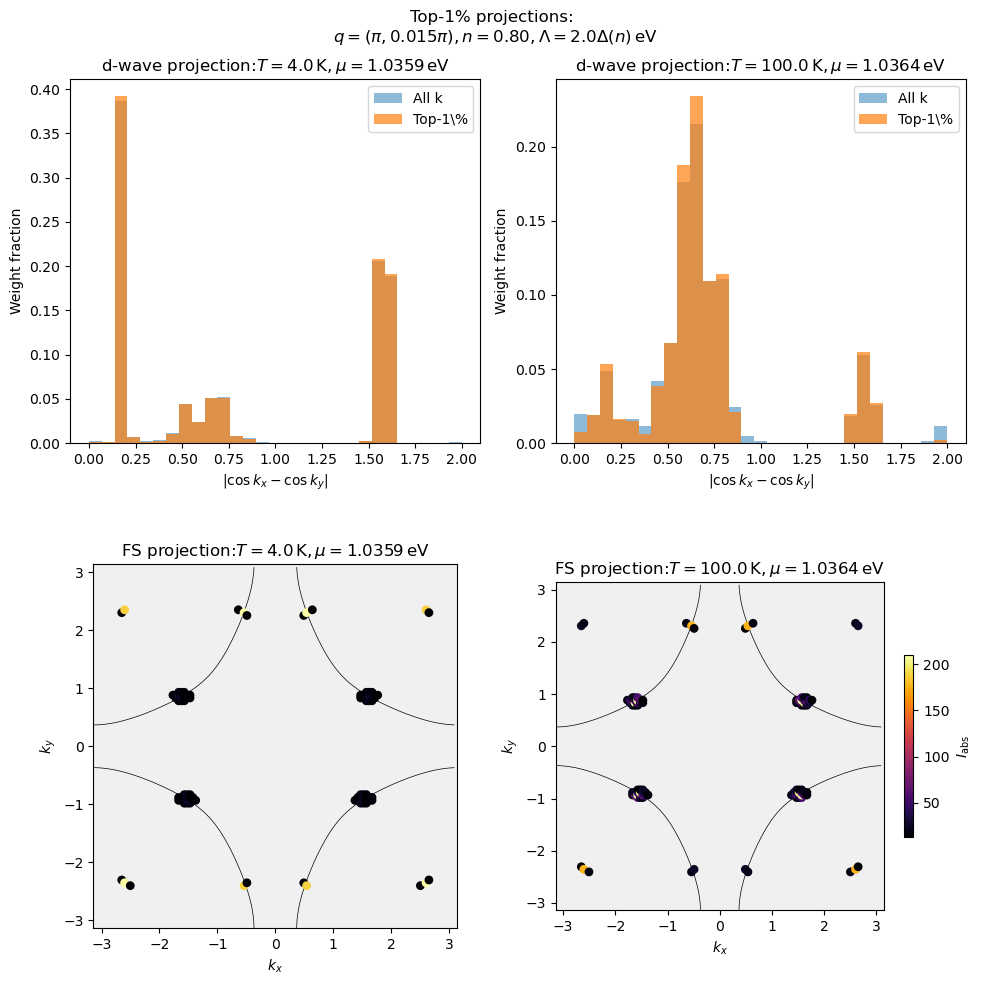}
    \caption{\footnotesize
    Top-1\% $I_{\mathbf{k}}(\mathbf{q})$ at $\mathbf{q}=(\pi,0i)$ 
    .}
    \label{fig:top1_pizero}
\end{subfigure}

\caption{\footnotesize
$\mathbf{k}$-resolved contributions to the static spin susceptibility intensity $I_{\mathbf{k};\mathbf{q}}$ and the corresponding top-1\% weighted distributions for $\mathbf{q}=(\pi,0)$.
(a) Brillouin-zone intensity maps at $n=0.8$ and $T=4$ K, which exhibits an apparent C2 symmetry with respect to $\mathbf{q}=(\pi, 0)$, with a dominant pair of scattering partners connected by a straight line.
(b) Corresponding top-1\% contributions projected onto the $d$-wave coordinate and the Fermi surface at $T=4$ K and $100$ K.
Parameters: $\Delta_{\mathrm{nom}}(n)=0.039$ eV and $\mu=1.0352$ eV for $n=0.8$. 
The ring-shaped scattering patterns around the M point are artifacts of the cumulant periodization procedure.
}
\label{fig:k_resolved_combined_pizero}
\end{figure*}

\begin{figure*}[t]
\centering

\begin{subfigure}{0.35\textwidth}
    \includegraphics[width=\linewidth]{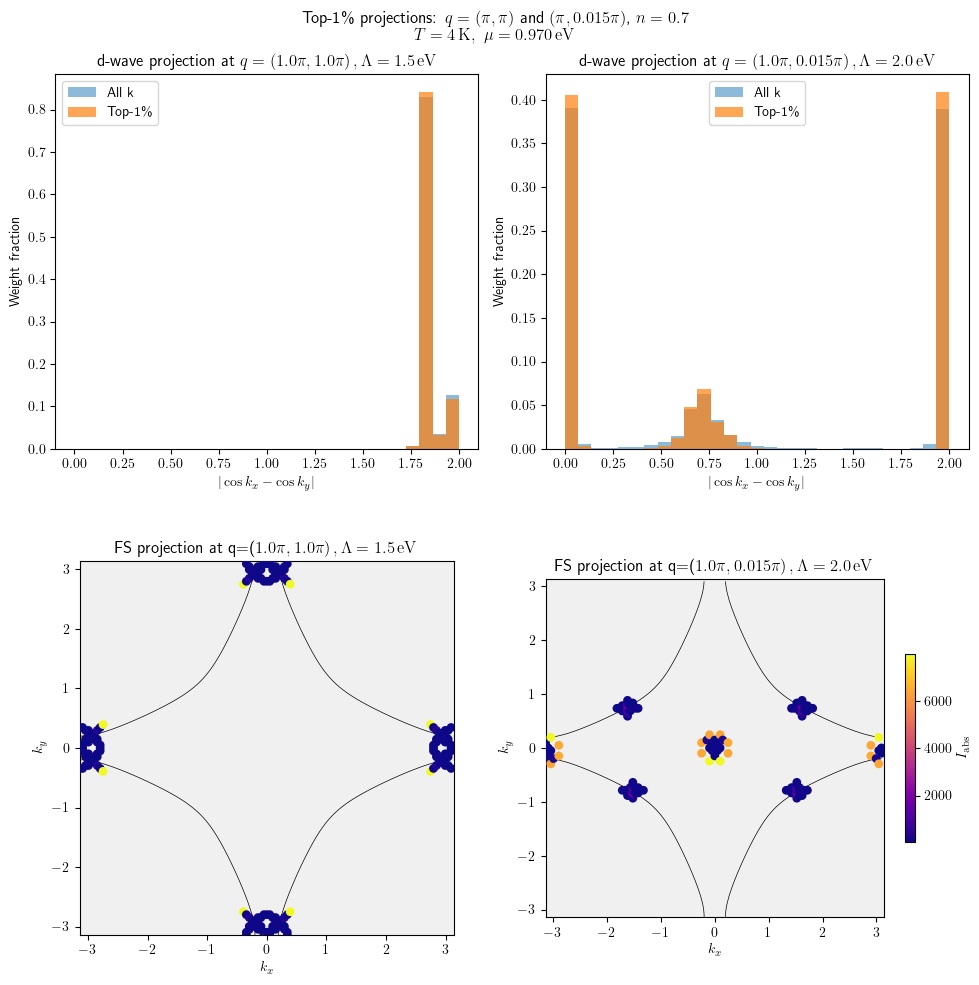}
    \caption{\footnotesize
   Top-1\% $I_{\mathbf{k}}(\mathbf{q})$ at $\mathbf{q}=(\pi,\pi)$ and $\mathbf{q}=(\pi,0)$ for $n=0.7$.}
    \label{fig:top1_vhs}
\end{subfigure}
\hfill
\begin{subfigure}{0.35\textwidth}
    \includegraphics[width=\linewidth]{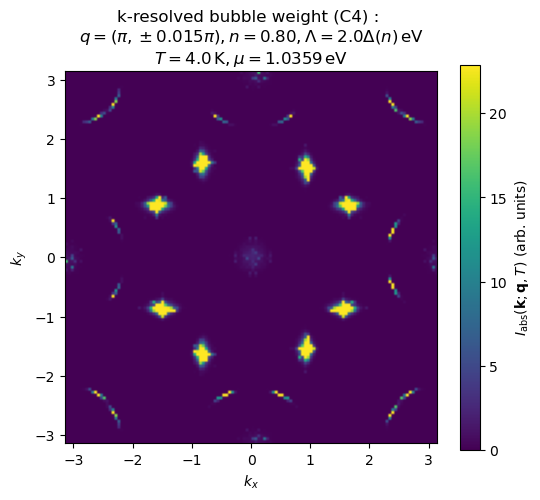}
    \caption{\footnotesize
    $I_{\mathbf{k}}(\mathbf{q})$ at $\mathbf{q}=(\pi,0)$ - C4.}
    \label{fig:bz_pizero_c4}
\end{subfigure}

\caption{\footnotesize
(a) $\mathbf{k}$-resolved top-1\% distribution in the VHS regime ($n=0.7$). Parameters: $\Delta_{\mathrm{nom}}(n)=0.051$ eV and  $\mu=0.970$ eV for $n=0.7$.
(b) Brillouin-zone intensity maps at $n=0.8$ and $T=4$ K for $\mathbf{q}=(\pi, \pm \delta)$, where the combined contributions restore $\mathbf{q}=(\pi, 0)$.
}
\label{fig:k_resolved_combined_others}
\end{figure*}

}
%%%%%%%%%%%%%%%%%%%%%%%%%%%%%%%%%%%%%%%%%%%%%%%%%%%%%%%%%%%%%%%%%%%%%%%%%%%%%%%%%%
%%%%%%%%%%%%%%%%%%%%%%%%%%%%%%%%%%%%%%%%%%%%%%%%%%%%%%%%%%%%%%%%%%%%%%%%%%%%%%%%%%

%%%%%%%%%%%%%%%%%%%%%%%%%%%%%%%%%%%%%%%%%%%%%%%%%%%%%%%%%%%%%%%%%%%%%%%%%%%%%%%%%%
%%%%%%%%%%%%%%%%%%%%%%%%%%%%%%%%%%%%%%%%%%%%%%%%%%%%%%%%%%%%%%%%%%%%%%%%%%%%%%%%%%
\section{Discussion}
{\small
% Uniform susceptibility
The uniform static spin susceptibility exhibits a Pauli-like temperature dependence with a weak quadratic ($T^{2}$) correction, consistent with an itinerant Fermi-liquid response characterized by a finite but modest quasiparticle density of states at $E_{F}$.
(Fig.~\ref{fig:chi_uniform_q0} and Table~\ref{tab:sommerfeld_q0}).

% thermal activated  features 
The static spin susceptibilities at $\mathbf{q}=(\pi,\pi)$ and $\mathbf{q}=(\pi,0)$ exhibit Arrhenius-type behavior over the range $n=0.8$--$0.95$, with $R^{2}>0.97$ and $R^{2}>0.91$, respectively.
At $\mathbf{q}=(\pi,0)$, $\Delta_{\mathrm{fit}}(n)<\Delta_{\mathrm{nom}}(n)$, whereas at $\mathbf{q}=(\pi,\pi)$, $\Delta_{\mathrm{fit}}(n)\sim\Delta_{\mathrm{nom}}(n)$, with consistently higher $R^{2}$ values. This behavior reflects the relatively stronger antinodal contribution and the reduced influence of low-energy nodal excitations 
(Figs.~\ref{fig:chi_static_pipi}\subref{fig:chi_pipi_main}--\ref{fig:chi_static_pizero}\subref{fig:chi_pizero_vhs} and Table~\ref{tab:static_arrhenius}).

% Onset temperatures of the spin susceptibility
Furthermore, the peak temperatures $T^\ast = T_{\mathrm{peak}}$ of the spin susceptibilities at finite $\mathbf{q}$ are comparable in magnitude to the superconducting transition temperatures $T_c$ in underdoped HTSCs at similar electron density $n$ (corresponding to hole doping $p = 1 - n$) ~\cite{reference10, reference11}.\\
At $\mathbf{q}=(\pi,\pi)$, 
\[
T^\ast = 33.7,\,65.9,\,107.9,\,100.5~\mathrm{K}, 
\]
at $\mathbf{q}=(\pi,0)$, 
\[
T^\ast = 9.2,\,40.7,\,84.9,\,101.7~\mathrm{K},
\]
for 
\[n = 0.95,\,0.9,\,0.85,\,0.8,
\]
corresponding to the antinodal gap amplitudes:
\[
\Delta_{\mathrm{nom}}(n)=0.010,\,0.021,\,0.034,\,0.039~\mathrm{eV},
\]
and the effective chemical potentials:
\[
\mu_{\mathrm{}}(4\mathrm{K}) =1.159,\,1.118,\,1.075,\,1.036~\mathrm{eV}, \mathrm{respectively},
\]
(Figs.~\ref{fig:chi_static_pipi}\subref{fig:chi_pipi_main}, \ref{fig:chi_static_pizero}\subref{fig:chi_pizero_main} and Table~\ref{tab:static_arrhenius}).
These values increase with increasing $\Delta_{\mathrm{nom}}(n)$, indicating that larger gap amplitudes shift the onset of thermally activated particle--hole excitations to higher temperatures.

% k-resolved contribution for Bz
The $\mathbf{k}$-resolved contributions indicate that although both channels contain contributions from nodal and antinodal regions, their dominant weights differ: for $\mathbf q=(\pi,\pi)$, the top-1\% $\mathbf{k}$-resolved contributions are concentrated near the antinodal regions, whereas for $\mathbf q=(\pi,0)$ they are distributed near near-nodal regions. (Figs.~\ref{fig:k_resolved_combined_pipi}\subref{fig:bz_pipi}--\ref{fig:k_resolved_combined_pizero}\subref{fig:top1_pizero}).

% k-resolved contribution for the Arrhenius behavior
The $\mathbf{k}$-resolved analysis further highlights that
the Arrhenius behavior of the spin susceptibility is consistent with the thermally activated emergence of antinodal contributions at both $\mathbf{q}=(\pi,\pi)$ and $\mathbf{q}=(\pi,0)$.
In both cases, antinodal intensity develops at $T = 100~\mathrm{K}$ but is strongly suppressed at $T = 4~\mathrm{K}$ for $n = 0.8$. (Figs.~\ref{fig:k_resolved_combined_pipi}\subref{fig:top1_pipi} and \ref{fig:k_resolved_combined_pizero}\subref{fig:top1_pizero}).

% k-resolved contribution for 
In addition, contributions from nodal regions and multiple momentum-transfer channels can reduce the effective activation scale (Eq.~\ref{eq:k_intensity}; Figs.~\ref{fig:k_resolved_combined_pipi}\subref{fig:top1_pipi} and \ref{fig:k_resolved_combined_pizero}\subref{fig:top1_pizero}; Table~\ref{tab:static_arrhenius}).
%For comparison, a momentum-independent $s$-wave gap generally produces a larger effective activation scale than a momentum-dependent $d$-wave gap because the latter retains low-energy nodal excitations.

% Chemical potential
The chemical potential $\mu$ used in this study is physically consistent, as it preserves the $(\mathbf{k},\mathbf{k}+\mathbf{q})$ phase space and samples the antinodal van Hove singularity (VHS) region.

It has a significant impact on the spin susceptibility at both $\mathbf{q}=(\pi,\pi)$ and $\mathbf{q}=(\pi,0)$.
As the chemical potential $\mu$ approaches the VHS regime, the spin susceptibility $\chi_{zz}$ is strongly enhanced in both channels, reflecting the increased density of states at the antinodes.

At $\mathbf{q}=(\pi,\pi)$, the spin susceptibility grows markedly at $n = 0.70$ with $\mu(4~\mathrm{K}) = 0.97~\mathrm{eV}$, while at $\mathbf{q}=(\pi,0)$ it increases significantly at $n = 0.60$ with $\mu(4~\mathrm{K}) = 0.925~\mathrm{eV}$, corresponding to regimes where the superconducting phase is absent in the phase diagram (Figs.~\ref{fig:chi_static_pipi}\subref{fig:chi_pipi_vhs} and \ref{fig:chi_static_pizero}\subref{fig:chi_pizero_vhs}).

The $\mathbf{k}$-resolved contributions further show that, in both cases, the influence of the antinodal VHS is clearly reflected in the intensity distribution around $n \approx 0.7$ (Fig.~\ref{fig:k_resolved_combined_others}\subref{fig:top1_vhs}).

% Self energy
The self-energy used in this study is physically consistent. It yields a stable chemical-potential offset, preserves a well-defined Fermi surface within the energy window relevant for $(\mathbf{k},\mathbf{k}+\mathbf{q})$ overlaps contributing to the spin susceptibility $\chi_{zz}$, and exhibits a Brillouin-zone average $\langle \mathrm{Re}\,\Sigma(i\omega_0)\rangle_{\mathrm{BZ}} = 1.51~\mathrm{eV}$, comparable to the Hartree self-energy of the same system in the paramagnetic state as shown in Appendix~B.%(Fig.~\ref{2024A_cdmft_fig1}). 
%The Gaussian energy-window scheme effectively isolates thermally-activated $\chi_{zz}$ in the system with this self-energy.

% Gaussian windows and extended Arrhenius fits
Other than that, the Gaussian energy-window, set to $\Lambda=1.25\,\Delta_{\mathrm{nom}}$ for all finite $\Delta$, yields thermally-activated susceptibility $\chi_{zz}$ effectively; the Arrhenius model in Eq.~(\ref{eq:fit2}) achieves $R^2>0.90$ for all finite $\Delta$.
However, the magnitude of the spin susceptibility $\chi_{zz}(\mathbf{q},T)$ obtained from the dressed-bubble is significantly reduced because the energy-window suppresses the available particle–hole phase space (Table~\ref{tab:static_arrhenius}).

From the pseudogap’s perspective, the pseudogap partially suppresses the antinodal spectral weight  $A(\mathbf{k}, \omega \approx 0)$,
while thermal activation or a chemical potential shift toward the van Hove level can restore low-energy particel-hole phase space there. 
The resulting enhancement of the spin susceptibility $\chi_{zz}$ correlates with reduced  $d_{x^2-y^2}$ pairing propensity, consistent with competition between axial-hole fluctuations and superconductivity.

The spin susceptibilities $\chi_{zz}$ at $\mathbf{q}=(\pi,\pi)$ and $\mathbf{q}=(\pi,0)$ are highly sensitive to the enhanced joint density of states associated with the antinodal van Hove saddle point at $(\pi,0)$.

}
%%%%%%%%%%%%%%%%%%%%%%%%%%%%%%%%%%%%%%%%%%%%%%%%%%%%%%%%%%%%%%%%%%%%%%%%%%%%%%%%%%
%%%%%%%%%%%%%%%%%%%%%%%%%%%%%%%%%%%%%%%%%%%%%%%%%%%%%%%%%%%%%%%%%%%%%%%%%%%%%%%%%%

%%%%%%%%%%%%%%%%%%%%%%%%%%%%%%%%%%%%%%%%%%%%%%%%%%%%%%%%%%%%%%%%%%%%%%%%%%%%%%%%%%
%%%%%%%%%%%%%%%%%%%%%%%%%%%%%%%%%%%%%%%%%%%%%%%%%%%%%%%%%%%%%%%%%%%%%%%%%%%%%%%%%%
\section{Conclusion}
{\small

We examine the temperature dependence of the momentum-resolved,
static spin susceptibility $\chi_{zz} (\mathbf{q}, \omega=0; T)$
in an underdoped Mott system with an antinodal gap, focusing on pseudogap-scale
low-energy excitations.

The spin susceptibilities for $\Delta > 0$ at the bond-direction (zone boundary) $\mathbf{q}=(\pi,\pi)$ and at the zone corner $\mathbf{q}=(\pi,0)$ show thermally-activated temperature dependences.
The spin susceptibility onset temperature tracks the critical temperature ($T_c$) of HTSCs with a comparable scale across the electron filling factor. That is, the superconductivity of HTSC disappears as the thermally-activated susceptibility appears.

Moreover, as the electron filling decreases and the chemical potential approaches the antinodal van Hove region, the electron-hole transitions at $\mathbf{q}=(\pi,\pi)$ and $\mathbf{q}=(\pi,0)$ grow markedly,
particularly in the regime where the superconducting phase is absent in the phase diagram.
%Additionally, the chemical potential approaching the van Hove region at antinodes may relate to
%the absence of a superconducting phase at the corresponding electron filling on the phase diagram.

Accordingly, it suggests that the emergence of cuprate superconductivity correlates with a suppression of low-energy antinodal spin response and associated particle-hole excitations,
which would otherwise dephase $d$-wave pairing, commonly attributed to spin fluctuations.
The pseudogap partially suppresses antinodal spectral weight near $\omega = 0$, thereby reducing the low-$\omega$ particle-hole phase space.
The static spin responses $\chi_{zz}(\mathbf{q}, \omega \to 0; T)$ at $\mathbf{q}=(\pi,\pi)$ and $\mathbf{q}=(\pi,0)$ effectively capture these antinodal effects.
%The pseudogap suppresses and redistributes spectral weight $A(\mathbf{k}, \omega)$ near $\omega=0$, especially at the antinodes, and suppresses the lifetimes of both spin fluctuations and particle-hole fluctuations. 

At last, we desire that this research will contribute to the fundamental elucidation of high-temperature oxide superconductivity.

}
%%%%%%%%%%%%%%%%%%%%%%%%%%%%%%%%%%%%%%%%%%%%%%%%%%%%%%%%%%%%%%%%%%%%%%%%%%%%%%%%%%
%%%%%%%%%%%%%%%%%%%%%%%%%%%%%%%%%%%%%%%%%%%%%%%%%%%%%%%%%%%%%%%%%%%%%%%%%%%%%%%%%%
%%%%%%%%%%%%%%%%%%%%%%%%%%%%%%%%%%%%%%%%%%%%%%%%%%%%%%%%%%%%%%%%%%%%%%%%%%%%%%%%%%
%%%%%%%%%%%%%%%%%%%%%%%%%%%%%%%%%%%%%%%%%%%%%%%%%%%%%%%%%%%%%%%%%%%%%%%%%%%%%%%%%%

%%%%%%%%%%%%%%%%%%%%%%%%%%%%%%%%%%%%%%%%%%%%%%%%%%%%%%%%%%%%%%%%%%%%%%%%%%%%%%%%%%
%%%%%%%%%%%%%%%%%%%%%%%%%%%%%%%%%%%%%%%%%%%%%%%%%%%%%%%%%%%%%%%%%%%%%%%%%%%%%%%%%%
%\usepackage{biblatex}
\bibliographystyle{unsrt}
\bibliography{ref_082526.bib}

\begin{thebibliography}{10}

\bibitem{reference1}
J.~Bardeen, L.~N. Cooper, and J.~R. Schrieffer.
\newblock Theory of superconductivity.
\newblock {\em Phys. Rev.}, 108:1175--1204, 1957.

\bibitem{reference2}
N.~F. Mott.
\newblock Metal-insulator transition.
\newblock {\em Rev. Mod. Phys.}, 40:677--683, 1968.

\bibitem{reference3}
J.~Hubbard.
\newblock Electron correlations in narrow energy bands.
\newblock {\em Proc. R. Soc. A}, 276:238--257, 1963.

\bibitem{reference4}
J.~G. Bednorz and K.~A. M{\"u}ller.
\newblock Possible high {$T_c$} superconductivity in the {Ba-La-Cu-O} system.
\newblock {\em Z. Phys. B}, 64:189--193, 1986.

\bibitem{reference5}
S.~R. White, D.~J. Scalapino, R.~L. Sugar, N.~E. Bickers, and R.~T. Scalettar.
\newblock Attractive and repulsive pairing interaction vertices for the
  two-dimensional {Hubbard} model.
\newblock {\em Phys. Rev. B}, 39:839--842, 1989.

\bibitem{reference6}
P~Monthoux, A.~V. Balatsky, and D.~Pines.
\newblock Towards a theory of high-temperature superconductivity in the
  antiferromagnetically correlated cuprate oxides.
\newblock {\em Phys. Rev. Letters}, 67:3448--3451, 1991.

\bibitem{reference7}
P.~I. Soininen, C.~Kallin, and A.~J. Berlinsky.
\newblock Structure of a vortex line in a $d_{x^2-y^2}$ superconductor.
\newblock {\em Phys. Rev. B}, 50:13883(R), 1994.

\bibitem{reference8}
C.~C. Tsuei, J.~R. Kirtley, C.~C. Chi, L.~S. Yu-Jahnes, A.~Gupta, T.~Shaw,
  A.~T. Fiory, and G.~P. Wildish.
\newblock Pairing symmetry and flux quantization in a tricrystal ring of
  {YBaCuO}.
\newblock {\em Phys. Rev. Lett.}, 73:593--596, 1994.

\bibitem{reference9}
M.~R. Norman, H.~Ding, M.~Randeria, J.~C. Campuzano, T.~Yokoya, T.~Takeuchi,
  T.~Takahashi, T.~Mochiku, P.~Guptasarma, K.~Kadowaki, and D.G. Hinks.
\newblock Destruction of the {Fermi} surface in underdoped high-{$T_c$}
  superconductors.
\newblock {\em Nature}, 392:157--160, 1998.

\bibitem{reference10}
W.~S. Lee, I.~M. Vishik, K.~Tanaka, D.~H. Lu, T.~Sasagawa, N.~Nagaosa, T.~P.
  Devereaux, Z.~Hussain, and Z.-X. Shen.
\newblock Abrupt onset of a second energy gap at the superconducting transition
  of underdoped {Bi2212}.
\newblock {\em Nature}, 450:81--84, 2007.

\bibitem{reference11}
Keishichiro Tanaka.
\newblock A study for the energy structure of the mott system with a low-energy
  excitation, in terms of the pseudo-gap in htsc.
\newblock {\em arXiv:2311.05865}, 2023.

\bibitem{reference12}
G.~Shirane, R.~J. Birgeneau, Y.~Endoh, P.~Gehring, M.~A. Kastner, K.~Kitazawa,
  H.~Kojima, I.~Tanaka, and T.~R. Thurston.
\newblock Temperature dependence of the magnetic excitations in
  {$\text{La}_{1.85}\text{Sr}_{0.15}\text{CuO}_{4}$} ({$T_{c}$}=33{K}).
\newblock {\em Phys. Rev. Lett.}, 63:330--333, 1989.

\bibitem{reference13}
Takashi Imai, Tadashi Shimizu, Hiroshi Yasuoka, Yutaka Ueda, and Koji Kosuge.
\newblock {NQR} and {NMR} studies of {$\text{YBa}_2\text{Cu}_3\text{O}_x$}
  ($6.0 \le x \le 6.91$).
\newblock {\em Int. J. Mod. Phys. B}, 3(1):93--95, 1989.

\bibitem{reference14}
T.~Dahm, V.~Hinkov, S.~V. Borisenko, A.~A. Kordyuk, V.~B. Zabolotnyy, J.~Fink,
  B.~B{\"u}chner, D.~J. Scalapino, W.~Hanke, and B.~Keimer.
\newblock Strength of the spin-fluctuation-mediated pairing interaction in a
  high-temperature superconductor.
\newblock {\em Nat. Phys.}, 5:217--221, 2009.

\bibitem{reference15}
L.~Wang, G.~He, Z.~Yang, et~al.
\newblock Paramagnons and high-temperature superconductivity in a model family
  of cuprates.
\newblock {\em Nat. Commun.}, 13:3163, 2022.

\bibitem{reference16}
J.~Lindhard.
\newblock On the properties of a gas of charged particles.
\newblock {\em K. Dan. Vidensk. Selsk. Mat.-Fys. Medd.}, 28(8):1--57, 1954.

\bibitem{reference17}
R.~Kubo.
\newblock Statistical-mechanical theory of irreversible processes. {I}.
  {General} theory of {Linear} response as applied to magnetic and electric
  fluctuations.
\newblock {\em J. Phys. Soc. Jpn.}, 12:570--586, 1957.

\bibitem{reference18}
G.~Kotliar, S.~Y. Savrasov, K.~Haule, V.~S. Oudovenko, O.~Parcollet, and C.~A.
  Marianetti.
\newblock Electronic structure calculations with dynamical mean-field theory.
\newblock {\em Rev. Mod. Phys.}, 78:865--951, 2006.

\bibitem{reference19}
A.~Georges, G.~Kotliar, W.~Krauth, and M.~J. Rozenberg.
\newblock Dynamical mean-field theory of strongly correlated fermion systems.
\newblock {\em Rev. Mod. Phys.}, 68:13--125, 1996.

\bibitem{reference20}
G.~D. Mahan.
\newblock {\em Many-Particle Physics}.
\newblock Kluwer Academic/Plenum, 3rd edition, 2000.

\bibitem{reference21}
P.~G. de~Gennes.
\newblock {\em Superconductivity of Metals and Alloys}.
\newblock W. A. Benjamin, 1966.

\bibitem{reference22}
Y.~Nambu.
\newblock Quasi-particles and gauge invariance in the theory of
  superconductivity.
\newblock {\em Phys. Rev.}, 117:648--663, 1960.

\bibitem{reference23}
H.~Park, K.~Haule, and G.~Kotliar.
\newblock Cluster dynamical mean field theory of the mott transition.
\newblock {\em Phys. Rev. Lett.}, 101:186403, 2008.

\bibitem{reference24}
E.~Gull, A.~J. Millis, A.~I. Lichtenstein, A.~N. Rubtsov, M.~Troyer, and
  P.~Werner.
\newblock Continuous-time monte carlo methods for quantum impurity models.
\newblock {\em Rev. Mod. Phys.}, 83:349--404, 2011.

\end{thebibliography}
%%%%%%%%%%%%%%%%%%%%%%%%%%%%%%%%%%%%%%%%%%%%%%%%%%%%%%%%%%%%%%%%%%%%%%%%%%%%%%%%%%
%%%%%%%%%%%%%%%%%%%%%%%%%%%%%%%%%%%%%%%%%%%%%%%%%%%%%%%%%%%%%%%%%%%%%%%%%%%%%%%%%%

%%%%%%%%%%%%%%%%%%%%%%%%%%%%%%%%%%%%%%%%%%%%%%%%%%%%%%%%%%%%%%%%%%%%%%%%%%%%%%%%%%
%%%%%%%%%%%%%%%%%%%%%%%%%%%%%%%%%%%%%%%%%%%%%%%%%%%%%%%%%%%%%%%%%%%%%%%%%%%%%%%%%%
\begin{acknowledgments}{\small The author acknowledges the use of ChatGPT (OpenAI) as a computational and language-assistance tool during the course of this work. All calculations, analyses, interpretations, and conclusions were performed and verified by the author.
}
%and Dr.Thomas Busch, Professor of Quantum Physics,
\end{acknowledgments}
%%%%%%%%%%%%%%%%%%%%%%%%%%%%%%%%%%%%%%%%%%%%%%%%%%%%%%%%%%%%%%%%%%%%%%%%%%%%%%%%%%
%%%%%%%%%%%%%%%%%%%%%%%%%%%%%%%%%%%%%%%%%%%%%%%%%%%%%%%%%%%%%%%%%%%%%%%%%%%%%%%%%%

%%%%%%%%%%%%%%%%%%%%%%%%%%%%%%%%%%%%%%%%%%%%%%%%%%%%%%%%%%%%%%%%%%%%%%%%%%%%%%%%%%
%%%%%%%%%%%%%%%%%%%%%%%%%%%%%%%%%%%%%%%%%%%%%%%%%%%%%%%%%%%%%%%%%%%%%%%%%%%%%%%%%%
\appendix

% APPENDIX A
%%%  APPENDIX A %%%%%%%%%%%%%%%%%%%%%%%%%%%%%%%%%%%%%%%%%%%%%%%%%%%%%%%%%%%%%%%%%%%%%%%%%%%%%%%
%%%%%%%%%%%%%%%%%%%%%%%%%%%%%%%%%%%%%%%%%%%%%%%%%%%%%%%%%%%%%%%%%%%%%%%%%%%%%%%%%%
\section{Related work}
\setcounter{equation}{0}
\renewcommand{\theequation}{A\arabic{equation}}

{\small

The results of a previous study using the Hubbard-1 approximation (Green's function methods), which motivated this study, are cited as reference ~\cite{reference11}.

The relationship between the electron density $n$ and the antinodal gap amplitude $\Delta (\equiv \Delta_{\mathrm{nom}})$ in this study follows the previous results in Table ~\ref{tab:previous_work}.
These results are in good agreement with the experimental data ~\cite{reference10}.

\begin{table}[H]
\centering
\caption{\footnotesize
Previously calculated values of the pseudogap \(\Delta (n)\) ($\mathrm{eV}$) at the antinodes of a cuprate model system are shown as a function of electron density (electron filling) $n$ ~\cite{reference10, reference11}. }
\label{tab:previous_work}
\begin{tabular}{lcccccc} 
  $n$ & 0.95 & 0.90 & 0.85 & 0.80 & 0.70 & 0.60 \\ 
%  $\Delta_{J} (\mathrm{eV})$ & 0.2375 & 0.225 & 0.2125 & 0.200 & 0.150\\
  $\Delta (n) (\mathrm{eV})$ & 0.010 & 0.021 & 0.034 & 0.039 & 0.051 & 0.054 \\
\end{tabular}
\end{table}

}
%%%%%%%%%%%%%%%%%%%%%%%%%%%%%%%%%%%%%%%%%%%%%%%%%%%%%%%%%%%%%%%%%%%%%%%%%%%%%%%%%%
%%%%%%%%%%%%%%%%%%%%%%%%%%%%%%%%%%%%%%%%%%%%%%%%%%%%%%%%%%%%%%%%%%%%%%%%%%%%%%%%%%

% APPENDIX B
%%%  APPENDIX B %%%%%%%%%%%%%%%%%%%%%%%%%%%%%%%%%%%%%%%%%%%%%%%%%%%%%%%%%%%%%%%%%%%%%%%%%%%%%%%
%%%%%%%%%%%%%%%%%%%%%%%%%%%%%%%%%%%%%%%%%%%%%%%%%%%%%%%%%%%%%%%%%%%%%%%%%%%%%%%%%%
\section{Cluster DMFT}
%%%%%%%%%%%%%%%%%%%%%%%%%%%%%%%%%%%%%%%%%%%%%%%%%%%%%%%%%%%%%%%%%%%%%%%%%%%%%%%%%%
%%%%%%%%%%%%%%%%%%%%%%%%%%%%%%%%%%%%%%%%%%%%%%%%%%%%%%%%%%%%%%%%%%%%%%%%%%%%%%%%%%
\setcounter{equation}{0}
\renewcommand{\theequation}{B\arabic{equation}}

\subsection{ CDMFT self-energy}
{\footnotesize
The self-energy is obtained using a 2$\times$2 plaquette Hubbard model and reconstructed into momentum space via cumulant periodization ~\cite{reference19, reference23}. 
The calculations are performed with the TRIQS CT-HYB impurity solver, 
using parameters \begin{math} U = 8.0\,t \end{math}, \begin{math} \beta = 100/t \end{math}, \begin{math} t^{'} =-0.3\,t \end{math}, and \begin{math} t=1.0 \end{math} (model units, corresponding to \begin{math} t=0.4~\mathrm{eV} \end{math}) ~\cite{reference24}.
The target filling is \begin{math} \rho =  0.8 \end{math} (\begin{math} n_{\mathrm imp} =  3.2 \end{math}; \begin{math} n_{\mathrm imp} =  4.0 \end{math} corresponds to the half-filling).
 A metallic solution is obtained by fixing the chemical potential at $\mu=3.2\,t$; the reference value at half-filling is determined from the particle–hole–symmetric limit, \begin{math}  \mu =  U/2  =  4.0\,t \end{math}. 
In the susceptibility analysis, all lattice energies are referenced to the chemical potential $\mu$ used in the tight-binding dispersion, so its absolute value cancels out.
The Brillouin-zone average of the real part of the self-energy at the lowest Matsubara frequency is $\langle \mathrm{Re}\,\Sigma(i\omega_0)\rangle_{\mathrm{BZ}} \approx 1.51~\mathrm{eV}$.

Fig.~\ref{2024A_cdmft_fig1} shows the mapping of the self-energy onto the high-symmetry points of the tight-binding model, exhibiting a moderate $\mathbf{k}$-dependence. 
The momentum dependence of $\mathrm{Re}\,\Sigma(\mathbf{k})$ is dominated by the nearest-neighbor harmonic $\gamma_{\mathbf{k}} = \tfrac{1}{2}(\cos k_{x} + \cos k_{y})$; 
the fitted positive coefficient yields a broad maximum near $\Gamma$.

\begin{figure}[tbp]
    \centering
    \includegraphics[width=0.35\textwidth]{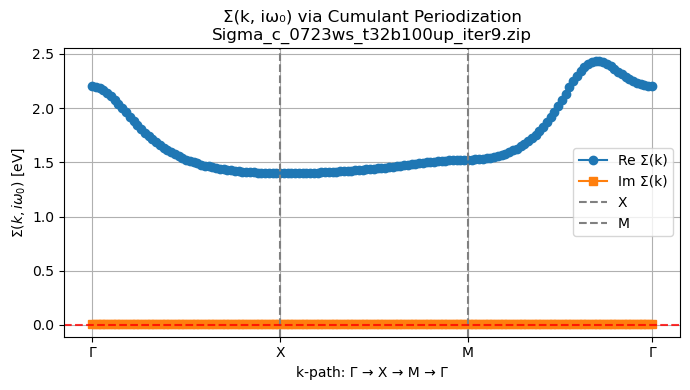}
     \caption{{\footnotesize
Self-energy along the high-symmetry path (in~$\mathrm{eV}$).
The Brillouin-zone-averaged value is 
$\langle \mathrm{Re}\,\Sigma(i\omega_0)\rangle_{\mathrm{BZ}} = 1.51~\mathrm{eV}$.
The upper curve shows $\mathrm{Re}\,\Sigma(\mathbf{k})$, and the lower curve shows $\mathrm{Im}\,\Sigma(\mathbf{k})$.
}} 
 \label{2024A_cdmft_fig1}
\end{figure}

\noindent\textit{Note.} The CDMFT loop in this study was terminated once the chemical potential reached the target occupancy; full self-consistency
$G_{\mathrm{imp}}(i\omega_n)\approx G_{\mathrm{loc}}(i\omega_n)$ was not enforced. Convergence near half-filling is quite
difficult because of the metal–insulator crossover. For reference, the Hartree self-energy in a paramagnet is $\sum^{H}_{}  = U n/ 2$. For $U=3.2~\mathrm{eV}$ and $n=0.80$, $\sum^{H} = 1.28~\mathrm{eV}$.
}
%%%%%%%%%%%%%%%%%%%%%%%%%%%%%%%%%%%%%%%%%%%%%%%%%%%%%%%%%%%%%%%%%%%%%%%%%%%%%%%%%%

%%%%%%%%%%%%%%%%%%%%%%%%%%%%%%%%%%%%%%%%%%%%%%%%%%%%%%%%%%%%%%%%%%%%%%%%%%%%%%%%%%
\subsection{CDMFT outline}
{\footnotesize
Cluster dynamical mean-field theory (CDMFT) solves a finite cluster embedded in a dynamical bath ~\cite{reference19, reference23}.

In the algorithm, the CT-HYB solve uses Monte Carlo sampling (hybridization expansion) with the Weiss Green’s function \begin{math} \mathit{g_{0}^{}} (i \omega_{n}) \end{math} (set by $H_{imp}$ and the hybridization $\Delta_{\mathrm{}}$),
and measures the impurity/cluster Green’s function \begin{math} G_{imp} (i \omega_{n}) \end{math}, and obtains the cluster self-energy \begin{math} \Sigma_{} (i \omega_{n}) \end{math} via the Dyson equation in (Eqs.~B1-2) ~\cite{reference24}.
%here $H_{imp}$ denotes the one-body part of impurity problem and $\Delta_{cdmft}$ is the hybridization function describing the coupling to the bath,

Both \begin{math} G_{loc} (i \omega_{n}) \end{math} and \begin{math} \Sigma_{} (i \omega_{n}) \end{math} are then used to construct a new Weiss Green's function via the Dyson equation in (Eq.~B5). 
The local Green’s function $G_{loc}$ is a coarse-grained version of the lattice Green’s function $G_{lattice} (\mathbf{k}, i\omega_{n} )$ by Brillouin-zone averaging over the cluster reduced Brillouin zone (Eqs.~B3-4). 

CDMFT iterates this self-consistency cycle until the convergence criterion is satisfied (Eq.~B6). 
At convergence, the resulting self-energy captures contributions from both the cluster and the bath, and is consistent with the coarse-grained lattice description.
A $\mathbf{k}$-dependent self-energy is then constructed after the solve by periodizing the cluster object to the lattice.\\

\paragraph{{\footnotesize CDMFT loop (Matsubara).}}
All quantities below are matrices in the cluster orbital (and spin) space.

\begin{align}
\intertext{Weiss Green’s function, representing the bath:}
g_{0}^{-1}(i\omega_{n}) &= i\omega_{n}\,\mathbf I + \mu\,\mathbf I - \Delta(i\omega_{n}) - H_{\mathrm{imp}},
\\[4pt]
\intertext{Impurity Dyson (TRIQS \texttt{solve()}):}
\Sigma(i\omega_{n}) &= g_{0}^{-1}(i\omega_{n}) - G_{\mathrm{imp}}^{-1}(i\omega_{n}),
\\[4pt]
\intertext{Lattice Green’s function at momentum $\mathbf{k}$ (reduced Brillouin zone):}
G_{\mathrm{latt}}(\mathbf{k}, i\omega_{n}) &= \Big[(i\omega_{n}+\mu)\,\mathbf I - t(\mathbf{k}) - \Sigma(i\omega_{n})\Big]^{-1},
\\[4pt]
\intertext{Coarse-grained (cluster-averaged) Green’s function:}
G_{\mathrm{loc}}(i\omega_{n}) &= \frac{1}{N_{\mathbf{k}}}\sum_{\mathbf{k}} G_{\mathrm{latt}}(\mathbf{k}, i\omega_{n}),
\\[4pt]
\intertext{Updated Weiss Green’s function for the next iteration:}
g_{0}^{-1}(i\omega_{n}) &= \Sigma(i\omega_{n}) + G_{\mathrm{loc}}^{-1}(i\omega_{n}),
\\[4pt]
\intertext{Self-consistency condition (at convergence):}
G_{\mathrm{imp}}(i\omega_{n}) &=
G_{\mathrm{loc}}(i\omega_{n}).
%\\[4pt]
%\intertext{Hybridization function:}
%\Delta(i\omega_{n}) &= (i\omega_{n}+\mu)\,\mathbf I - H_{\mathrm{imp}} - g_{0}^{-1}(i\omega_{n}).
\end{align}
}
%%%%%%%%%%%%%%%%%%%%%%%%%%%%%%%%%%%%%%%%%%%%%%%%%%%%%%%%%%%%%%%%%%%%%%%%%%%%%%%%%%
%%%%%%%%%%%%%%%%%%%%%%%%%%%%%%%%%%%%%%%%%%%%%%%%%%%%%%%%%%%%%%%%%%%%%%%%%%%%%%%%%%

% APPENDIX C
%%%%%%%%%%%%%%%%%%%%%%%%%%%%%%%%%%%%%%%%%%%%%%%%%%%%%%%%%%%%%%%%%%%%%%%%%%%%%%%%%%
%%%%%%%%%%%%%%%%%%%%%%%%%%%%%%%%%%%%%%%%%%%%%%%%%%%%%%%%%%%%%%%%%%%%%%%%%%%%%%%%%%
\section{Linear-response derivation of the Lindhard expression}
\label{app:lindhard}
\setcounter{equation}{0}
\renewcommand{\theequation}{C\arabic{equation}}

{\footnotesize
Consider a time-dependent perturbation \(H'(t)=-f(t)\,A\) that couples an external field \(f(t)\) to the operator \(A\), yielding a change in an observable \(B\).
The retarded (real time) susceptibility (Kubo formula) is given by: ~\cite{reference16, reference17, reference18, reference19}.

\begin{align}
\chi^{R}_{AB}(t) = -\,i\,\theta(t)\,\langle [A(t),B(0)]\rangle .
\end{align}

On imaginary time (\(\tau\)) axis, 
\begin{align}
\chi_{AB}(\tau) &= -\,\langle T_{\tau}\,A(\tau)\,B(0)\rangle^{\mathrm{conn}}, \\
\chi_{AB}(i\Omega_m) &= \int_{0}^{\beta} d\tau\, e^{\,i\Omega_m \tau}\, \chi_{AB}(\tau),
\end{align}
where \(T_{\tau}\) is the imaginary-time ordering operator and \(\Omega_m=2\pi m T\) are bosonic Matsubara frequencies.

For density response, taking \(A=B=\rho_{\mathbf{q}}\) witth:
\begin{align}
\rho_{\mathbf{q}}=\sum_{\mathbf{k},\sigma} c^{\dagger}_{\mathbf{k}+\mathbf{q},\sigma}\,c_{\mathbf{k},\sigma},
\end{align}
we obtain:
\begin{align}
\chi_{0}(\mathbf{q},\tau)= -\,\big\langle T_{\tau}\,\rho_{\mathbf{q}}(\tau)\,\rho_{-\mathbf{q}}(0)\big\rangle^{\mathrm{conn}}.
\end{align}

Through Wick’s theorem for the non-interacting average and using the Green's function representation:
\begin{align}
G_{0}(\mathbf{k},\tau)= -\,\big\langle T_{\tau}\,c_{\mathbf{k}}(\tau)\,c^{\dagger}_{\mathbf{k}}(0)\big\rangle_{0},
\end{align}
the connected contraction (survival) gives the bubble, where the subscript 0 denotes the non-interacting (reference) thermal average:
\begin{align}
\chi_{0}(\mathbf{q},\tau)= -\sum_{\mathbf{k},\sigma} G_{0}(\mathbf{k},\tau)\,G_{0}(\mathbf{k}+\mathbf{q},-\tau).
\end{align}

Fourier transforming in imaginary time (bosonic $\Omega_m$) via \allowbreak
Eq.~$\big(C2\big)-\big(C3\big)$ and using the $\lbrack0,\beta \rbrack$ normalization,
while using the $\tau$-integral to enforce $\nu'_n=\nu_n+\Omega_m$, yields:
\begin{align}
\chi_{0}(\mathbf{q},i\Omega_m)
= -(1/\beta) \sum_{\mathbf{k},\nu_n} G_{0}(\mathbf{k},i\nu_n)\,
G_{0}(\mathbf{k}+\mathbf{q},i\nu_n+i\Omega_m),
\end{align}
where \(\nu_n=(2n+1)\pi T\) are fermionic Matsubara frequencies.
The external bosonic frequency is the transfer in the Feynman diagram, so it appears as the difference between the two fermionic loop frequencies; 
by shifting the fermionic Matsubara index $\nu_{n}$, one can place $i \Omega_{m}$ on either line, but not both.

After performing the fermionic Matsubara sum, analytic continuation
\(i\Omega_m \to \omega+i0^{+}\) yields:
\begin{align}
\chi_{0}^{zz}(\mathbf{q},\omega)
= -\,\frac{(g\mu_B)^2}{N}\sum_{\mathbf{k}}
\frac{f(\varepsilon_{\mathbf{k}})-f(\varepsilon_{\mathbf{k}+\mathbf{q}})}
{\hbar\omega+\varepsilon_{\mathbf{k}}-\varepsilon_{\mathbf{k}+\mathbf{q}}+i0^{+}},
\end{align}

and for the transverse (spin-flip) response:
\begin{align}
\chi_{0}^{+-}(\mathbf{q},\omega)
&= -\,\frac{(g\mu_B)^2}{N}\sum_{\mathbf{k}}
\frac{f(\varepsilon_{\mathbf{k},\uparrow})-f(\varepsilon_{\mathbf{k}+\mathbf{q},\downarrow})}
{\hbar\omega+\varepsilon_{\mathbf{k},\uparrow}-\varepsilon_{\mathbf{k}+\mathbf{q},\downarrow}+i0^{+}}.
\end{align}
In a paramagnet without Zeeman splitting, $\varepsilon_{\mathbf{k},\uparrow}=\varepsilon_{\mathbf{k},\downarrow}$.\\

Note that the connected part isolates fluctuations:
\begin{align}
C^{\mathrm{conn}}_{AB}(\tau)
\equiv \langle T_{\tau} A(\tau) B(0) \rangle - \langle A \rangle \langle B \rangle ,
\end{align}
where \(\langle \cdots \rangle\) denotes the equilibrium (thermal) average.

Assuming SU(2) spin-rotational symmetry and zero field, we denote the bare longitudinal spin susceptibility by
$\chi_{0} (\mathbf q,\omega)\equiv\chi_{0}^{zz}(\mathbf q,\omega)=\tfrac12\,\chi_{0}^{+-}(\mathbf q,\omega)$, 
where $S^{\pm}=S^{x}\pm iS^{y}$.
}
%%%%%%%%%%%%%%%%%%%%%%%%%%%%%%%%%%%%%%%%%%%%%%%%%%%%%%%%%%%%%%%%%%%%%%%%%%%%%%%%%%
%%%%%%%%%%%%%%%%%%%%%%%%%%%%%%%%%%%%%%%%%%%%%%%%%%%%%%%%%%%%%%%%%%%%%%%%%%%%%%%%%%

\end{document}